\documentclass[english,twoside]{article}
\usepackage[latin9]{inputenc}
\usepackage{array}
\usepackage{booktabs}
\usepackage{amsmath}
\usepackage{amsthm}
\usepackage{amssymb}
\usepackage{graphicx} 
\usepackage[normalem]{ulem}
\usepackage{xcolor}
\usepackage{placeins}
\newtheorem{prop}{Proposition}

\makeatletter

\definecolor{myb}{RGB}{0, 102, 255}

\providecommand{\tabularnewline}{\\}

\numberwithin{equation}{section}
\numberwithin{figure}{section}
\def\R{{\mathbb R}}
\def\cl#1{{\cal #1}}
\def\E{{\mathbb E}}

\usepackage{a4wide}
\usepackage{babel}
\usepackage{fullpage}
\usepackage{authblk}
\usepackage{hyperref}
\usepackage{placeins}
\usepackage{hyphenat} 
\usepackage{fancyvrb}
\usepackage[hypcap]{caption}
\usepackage{xcolor}
\usepackage{amsthm}
\usepackage{float}

\newtheorem{remark}{Remark}
\title{Variance Reduction Applied to Machine Learning for Pricing Bermudan/American Options in High Dimension}
\author{ \textsc{Ludovic Gouden\`ege} \thanks{F\'ed\'eration de Math\'ematiques de CentraleSup\'elec - CNRS FR3487, France -\texttt{ ludovic.goudenege@math.cnrs.fr}} \and \textsc{Andrea Molent} \thanks{Dipartimento di Scienze Economiche e Statistiche, Universit\`a degli Studi di Udine, Italy - \texttt{andrea.molent@uniud.it}} \and \textsc{Antonino Zanette}\thanks{Dipartimento di Scienze Economiche e Statistiche, Universit\`a degli Studi di Udine, Italy - \texttt{antonino.zanette@uniud.it}} \footnote{Corresponding author.}}
\date{}

\makeatother

\usepackage{babel}
\begin{document}
	\maketitle

	\begin{flushleft}
		\rule{1\columnwidth}{1pt}
		\par\end{flushleft}
	
	\begin{flushleft}
		\textbf{\large{}Abstract}
		\par\end{flushleft}{\large \par} 	
In this paper we propose an efficient method to compute the price of multi-asset American  options, based on Machine Learning, Monte Carlo simulations and variance reduction technique. Specifically, the options we consider are written on a basket of assets, each of them following a Black-Scholes dynamics.
 In the wake of Ludkovski's approach \cite{ludkovski2018Kriging}, we implement here a backward dynamic programming  algorithm which considers a finite number of uniformly distributed exercise dates.  On these dates, the option value is computed as the maximum between the exercise value and the continuation value, which is obtained by means of Gaussian process regression technique and Monte Carlo simulations. 
Such a method performs well for low dimension baskets but it is not accurate for very high dimension baskets.  In order to improve the dimension range, we employ the European option price as a control variate, which allows us to treat very large baskets and moreover to reduce the variance of  price estimators.
Numerical tests show that the proposed algorithm is fast and reliable, and it can handle also American options on very large baskets of assets, overcoming the problem of the curse of dimensionality. 
	\vspace{2mm}
	
	\noindent \emph{\large{}Keywords:} \\ Finance; Gaussian process regression; Control  variate; American options; Monte Carlo methods.\\
	\noindent\rule{1\columnwidth}{1pt}
	\newpage
	\section{Introduction}
	
In this paper we consider one of the most compelling problems among the still open issues in the field of  computational finance: pricing and hedging American options in high dimension. 
From a practical point of view, the efficient numerical evaluation of American options which consider as underlying a baskets of $d$ assets is very challenging because of the so-called ``curse of dimensionality'', which avoids the direct application of standard numerical schemes such as finite difference or tree methods. Specifically, this curse of dimensionality means that the computational cost and the memory requirement  increase exponentially with the dimension of the problem.\\
Several new ideas have appeared in this research area, which can be divided into five groups. The first type of approach consists in employing a recombinant tree in order to obtain a discretization of the underlying diffusion. An example of this mode is given by
the stochastic mesh method of Broadie and Glasserman \cite{broadie1997pricing}, the quantization
algorithms of Bally, et al.
\cite{bally2003first}, the stochastic grid method of Jain and Oosterlee \cite{jain2012pricing}. 
The second idea makes use of regression on a truncated basis of $L^{2}$ in order to compute the
conditional expectations. This is done in Longstaff and Schwartz
\cite{longstaff2001valuing} and in
Tsisiklis and Van Roy \cite{tsitsiklis1999optimal}. 
The third concept consists in exploiting the representation formulas for the conditional expectation using Malliavin calculus.
This has been done by Lions and Reigner \cite{lions2001calcul}, Bouchard and Touzi \cite{bouchard2004discrete}, Bally et al. \cite{bally2005pricing} Caramellino and Zanette \cite{caramellino2011monte} and Abbas-Turki and Lapeyre\cite{abbas2012american}. 
Another group of ideas relies on duality-based approaches for Bermudan option pricing, which are proposed by Rogers \cite{rogers2002monte}, Haugh and Kogan \cite{haugh2004pricing},  Andersen and Broadie \cite{andersen2004primal}, Schoenmakers et al. \cite{schoenmakers2013optimal} and Lelong \cite{lelong2018dual}, which can be used to construct bounds on the option value.
Finally, the last group consists of methods that employ Machine Learning techniques to learn the continuation value or the stopping rules. This has been proposed by Becker et al. \cite{becker2019deep}, Kohler et al. \cite{kohler2010pricing} and Ludkowski \cite{ludkovski2018Kriging}.

European prices can be used as control variate while pricing American options, as done, for example, by Bally et al. \cite{bally2005pricing} and by Caramellino and Zanette \cite{caramellino2011monte}.  Since multi-asset products are considered, efficiently computing European prices is not trivial and many authors developed valid methods in this field.  
	Some of them focused on computing lower and upper bounds, such as Deelstra et al. \cite{deelstra2004pricing},   Carmona and Durrleman \cite{carmona2005generalizing}, Caldana et al. \cite{caldana2016general}.
	Other approaches for basket options are based on the approximation of the sum of the log-normal distributions with a simple  distribution by matching some moments, as done by Levy \cite{levy1992pricing}, Milevsky and Posner \cite{milevsky1998closed,posner1998valuing}, Zhou and Wang \cite{zhou2008accurate}, Korn and Zeytun \cite{korn2013efficient}. Moreover, an approximation approach is also proposed  by Li and Wu \cite{li2008approximation} for options on several mean-reverting assets.
	Recently, Glau et al. \cite{glau2019low} and Glau et al. \cite{glau2019new} consider Chebyshev based methods for pricing. Deep Learning techniques are nowadays widely used in solving large differential equations, which is intimately related to option pricing: recent progresses in this field have been achieved by  Han et al. \cite{han2018solving}, E et al. \cite{e2017deep} and Beck et al.  \cite{beck2019Machine}.	Finally, efficient Monte Carlo approaches are developed by Jourdain and Lelong \cite{jourdain2009robust} and more recently by Bayer et al. \cite{bayer2018smoothing}.

In this paper, we propose a new method that combines  Machine Learning, Monte Carlo simulations and variance reduction control variate technique. In particular, the use of a control variate makes the method more stable and extends its applicability range to high very large baskets. Moreover, the variance of price estimator is significantly reduced.

First of all, we implement a version  of the Ludkovski's algorithm \cite{ludkovski2018Kriging}. Such an algorithm  proceeds backward over time by computing the price function on a set of prearranged points which represents possible values of the underlying. In particular, at each time step, it uses a set of Monte Carlo simulations together with Gaussian Process Regression (GPR)  to approximate the continuation value at these points. The  option price is then obtained as the maximum between the continuation value and the intrinsic value of the option. We term such an algorithm GPR Monte Carlo (GPR-MC).
The GPR-MC algorithm works very well for small baskets (in his paper, Ludkovski considers up to 5 dimensional basket), but it does not for large ones. In this paper, we show that, if one considers the European price as a control variate, the algorithm improves significantly and the variance of the price estimator is reduced. We term GPR Monte Carlo Control Variate (GPR-MC-CV) this new algorithm. Moreover, in order to compute the European prices, we suggest to use a semi-analytical formula, named  GPR-EI formula, introduced by Gouden\`ege et al. in \cite{goudenege2019machine}, which proves to be efficient when many repeated computations of European prices have to be performed, or alternatively, Quasi-Monte Carlo simulations. 
 Finally, we investigate   the benefits brought by control variate technique to the GPR-Tree and GPR-EI approaches introduced by Gouden\`ege et al. \cite{goudenege2019machine}.
	The paper is organized as follows. In Section 2 we
	present American options for the Black-Scholes $d$-dimensional model. In Section
	3 we briefly review Gaussian Process Regression, we present the GPR-EI formula, the GPR-MC method and the GPR-MC-CV method. Furthermore, we also investigate the use of control variate technique for the GPR-Tree and GPR-EI methods. In Section 4 we report
    some numerical results about pricing and variance reduction. Finally, Section 5 draws the conclusions.

	\section{American options in the multi-dimensional Black-Scholes model}
An American option  with maturity $T$ is a derivative instrument whose holder can exercise the intrinsic optionality at any moment, from inception up to maturity.
Let $\mathbf{S}=(\mathbf{S}_t)_{t\in [0,T]}$ denote the $d$-dimensional underlying  process.
Such a stochastic process is assumed to randomly evolve according to the multidimensional Black-Scholes model: under the risk neutral measure, such a model is given by the following equation
\begin{equation}\label{sde}
dS^i_t=\left(r-\eta_{i} \right) \, S^i_t\, dt +\sigma_{i}\, S^i_t\, dW^i_t,\quad
\  i=1,\ldots, d,  
\end{equation}
with $\mathbf{S}_0=\left( s_{0,1},\dots,s_{0,d}\right)^{\top} \in \R_+^d$ the spot price, $r$  the (constant) spot interest rate, $\boldsymbol{\eta}=(\eta_1, \dots,
\eta_d)^{\top}$ the vector of (constant) dividend rates, 
 $\boldsymbol{\sigma} = \left( \sigma_1, \dots,\sigma_d\right)^{\top} $  the vector of (constant) volatilities, $\mathbf{W}$ a $d$-dimensional correlated Brownian motion and $\rho_{ij}$ the
instantaneous correlation coefficient between $W^i_t$ and $W^j_t.$
Moreover, let $\Psi(\mathbf{S}_T)$ denote the cash-flow associated with the option.
Thus, the price at time $t$ of an American option having maturity $T$
and payoff function $\Psi\, :\, \R_+^d\to \R$ is then
\begin{equation}\label{price}
v^{AM}(t,\mathbf{x})=\sup_{\tau\in \cl T_{t,T}}
\E_{t,\mathbf{x}}\left[ e^{-r (\tau-t) }
\Psi(\mathbf{S}_\tau)\right],
\end{equation}
	 where $\cl T_{t,T}$ stands for the set of all the stopping times
taking values on $[t,T]$ and $\E_{t,\mathbf{x}}\left[ \cdot \right]$ is the expectation given all the information at time $t$ and assuming $\mathbf{S}_t=\mathbf{x}$.

For simulation purposes, the $d-$dimensional Black-Scholes model can be written alternatively using the Cholesky decomposition. 
Specifically, for $i \in \{1, \dots, d\}$ we can write
\begin{equation}\label{sde_cho}
  dS^i_t = S^i_t (\left(r-\eta_{i} \right)  dt + \sigma_i \Sigma_i d\mathbf{B}_t),   
\end{equation}
where $\mathbf{B}$ is a d-dimensional Brownian motion and $\Sigma_i$ is the $i$-th row of the matrix $\Sigma$ defined as a square root of the
correlation matrix $\Gamma$, given by
\begin{equation}
  \Gamma = \begin{pmatrix}
    1 & \rho_{12} & \hdots & \rho_{1d}\\
    \rho_{21} & 1 &\ddots & \vdots\\
    \vdots&\ddots&\ddots& \vdots\\
    \rho_{d1} &\hdots  &\hdots & 1 
  \end{pmatrix}.
\end{equation}
	\section{Machine Learning for American options in the multi-dimensional Black-Scholes model}
	
	\subsection{Gaussian Process Regression}
	
	In this Section, we present a brief review of Gaussian Process Regression
	and for a comprehensive treatment we refer to Rasmussen and Williams
	\cite{rasmussen2006gaussian}.
	
	Gaussian Process Regression (GPR),  also known as Kriging (see Matheron \cite{matheron1973intrinsic}, Journel and Huijbregts \cite{journel1978mining}), is a class of non-parametric
	kernel-based probabilistic models which represents the input data
	as the random observations of a Gaussian stochastic process. The most
	important advantage of this approach in relation to other parametric
	regression techniques is that it is possible to effectively exploit
	a complex dataset which may consist of points sampled randomly in
	a multidimensional space.
	
	In general, a Gaussian process $\mathcal{G}$ is a collection of random
	variables defined on a common probability space ${\displaystyle (\Omega,\mathcal{F},P)}$,
	any finite number of which have consistent joint Gaussian distributions.
	We are interested in Gaussian processes for which the random variables
	in $\mathcal{G}$ are indexed by a point $\mathbf{x\in}\mathbb{R}^{d}$,
	$d\in\mathbb{N}$. Therefore, for all $\mathbf{x}\in\mathbb{R}^{d}$,
	$\mathcal{G}\left(\mathbf{x}\right):\Omega\rightarrow\mathbb{R}$
	is a Gaussian random variable and if $X$=$\left\{ \mathbf{x}_{p},p=1,\dots,P\right\} \subset\mathbb{R}^{d}$
	then $\left(\mathcal{G}\left(\mathbf{x}_{1}\right),\dots,\mathcal{G}\left(\mathbf{x}_{P}\right)\right)^{\top}$
	is a random Gaussian vector. Moreover, a Gaussian process is fully
	specified by its mean function $\mu\left(\mathbf{x}\right):\mathbb{R}^{d}\rightarrow\mathbb{R}$
	(which is usually assumed to be zero) and by its covariance function $k\left(\mathbf{x},\mathbf{x}'\right):\mathbb{R}^{d}\times\mathbb{R}^{d}\rightarrow\mathbb{R}$. 
	
	Now, let us consider a training set $\mathcal{D}$ of $P$ observations
	(the input data), $\mathcal{D}=\left\{ \left(\mathbf{x}_{p},y_{p}\right), p=1,\dots,P\right\} $
	where $X=\left\{ \mathbf{x}_{p},p=1,\dots,P\right\} \subset\mathbb{R}^{d}$
	denotes the set of input vectors and $Y=\left\{ y_{p},p=1,\dots,P\right\} \subset\mathbb{R}$
	denotes the set of scalar outputs. These observations are modeled
	as the realization of the sum of a Gaussian process and a noise source.
	Specifically,
	\begin{equation}
	y_{p}=f_{p}+\varepsilon_{p},
	\end{equation}
	where $\left\{ f_{p}=\mathcal{G}\left(\mathbf{x}_{p}\right),p=1,\dots,P\right\} $
	is a Gaussian process and $\left\{ \varepsilon_{p},p=1,\dots,P\right\} $
	are i.i.d. random variables such that $\varepsilon_{p}\sim\mathcal{N}\left(0,\sigma_{P}^{2}\right)$.
	Moreover, the distribution of $\mathbf{f}=\left(f_{1}\dots f_{P}\right)^{\top}$
	is assumed to be given by
	\begin{equation}
	\mathbf{f}\sim\mathcal{N}\left(\mathbf{0},K\left(X,X\right)\right),
	\end{equation}
	where $K\left(X,X\right)$ is a $P\times P$ matrix with $K\left(X,X\right)_{p_1,p_2}=k\left(\mathbf{x}_{p_1},\mathbf{x}_{p_2}\right)$ for $p_1,p_2=1,\dots,P$ with $k:\mathbb{R}^d\times \mathbb{R}^d\rightarrow \mathbb{R}$   the so called   kernel function.
	Thus
	\begin{equation}
	\mathbf{y}\sim\mathcal{N}\left(\mathbf{0},K\left(X,X\right)+\sigma_{P}^{2}I_{P}\right),
	\end{equation}
	where $I_{P}$ is the $P\times P$ identity matrix.
	
	Now, in addition, let us consider a test set $\tilde{X}$ of $M$
	points $\left\{ \tilde{\mathbf{x}}_{m},m=1,\dots,M\right\} $. The
	realizations $\tilde{f}_{m}=\mathcal{G}\left(\tilde{\mathbf{x}}_{m}\right)$
	are not known but rather we want to estimate them by exploiting the
	observed realizations of $\mathcal{G}$ in $\mathcal{D}$. The \emph{a
		priori} joint distribution of $\mathbf{y}$ and $\mathbf{\tilde{f}}=\left(\tilde{f}_{1},\dots,\tilde{f}_{M}\right)^{\top}$
	is given by 
	\begin{equation}
	\left[\begin{array}{c}
	\mathbf{y}\\
	\mathbf{\tilde{f}}
	\end{array}\right]\sim\mathcal{N}\left(\left[\begin{array}{c}
	\mathbf{0}_{P}\\
	\mathbf{0}_{M}
	\end{array}\right],\left[\begin{array}{cc}
	K\left(X,X\right)+\sigma_{P}^{2}I_{P} & K\left(X,\tilde{X}\right)\\
	K\left(\tilde{X},X\right) & K\left(\tilde{X},\tilde{X}\right)
	\end{array}\right]\right)
	\end{equation}
	where $K\left(\tilde{X},\tilde{X}\right)$ is a $M\times M$ matrix
	given by $K\left(\tilde{X},\tilde{X}\right)_{m_1,m_2}=k\left(\mathbf{\tilde{x}}_{m_1},\mathbf{\tilde{x}}_{m_2}\right)$ for $m_1,m_2=1,\dots,M$,
	$K\left(X,\tilde{X}\right)$ is a $P\times M$ matrix given by $K\left(X,\tilde{X}\right)_{p,m}=k\left(\mathbf{x}_{p},\mathbf{\tilde{x}}_{m}\right)$ for $p=1,\dots,P$, $m=1,\dots,M$
	and $K\left(\tilde{X},X\right)$ is a $M\times P$ matrix given by
	$K\left(\tilde{X},X\right)_{m,p}=k\left(\mathbf{\tilde{x}}_{m},\mathbf{x}_{p}\right)$ for $m=1,\dots,M$, $p=1,\dots,P$. 
	
	Since we know the values for the training set, we can consider the
	conditional distribution of $\mathbf{\tilde{f}}$ given $\mathbf{y}$.
	It is possible to prove that $\mathbf{\tilde{f}}|\tilde{X},\mathbf{y},X$
	follows a Gaussian distribution given by 
	\begin{equation}
	\mathbf{\tilde{f}}|\tilde{X},\mathbf{y},X\sim\mathcal{N}\left(\mathbb{E}\left[\mathbf{\tilde{f}}|\tilde{X},\mathbf{y},X\right],Cov\left[\mathbf{\tilde{f}}|\tilde{X},\mathbf{y},X\right]\right),
	\end{equation}
	where 
	\begin{equation}
	\mathbb{E}\left[\mathbf{\tilde{f}}|\tilde{X},\mathbf{y},X\right]=K\left(\tilde{X},X\right)\left[K\left(X,X\right)+\sigma_{P}^{2}I_{P}\right]^{-1}\mathbf{y}\label{eq:prediction}
	\end{equation}
	and
	\begin{equation}
	Cov\left[\mathbf{\tilde{f}}|\tilde{X},\mathbf{y},X\right]=K\left(\tilde{X},\tilde{X}\right)-K\left(\tilde{X},X\right)\left[K\left(X,X\right)+\sigma_{P}^{2}I_{P}\right]^{-1}K\left(X,\tilde{X}\right).
	\end{equation}
	Therefore, a natural choice consists in predicting the values $\mathbf{\tilde{f}}$
	through $\mathbb{E}\left[\mathbf{\tilde{f}}|\tilde{X},\mathbf{y},X\right]$. 
	Moreover, by using equation  \eqref{eq:prediction},  one can define a function $f^{GPR}:\mathbb{R}^d\rightarrow \mathbb{R}$ that approximates the function $\mathbf{x}_p\mapsto y_p$ by setting
	\begin{align}
	f^{GPR}\left( \tilde{\mathbf{x}}\right) &=\mathbb{E}\left[\mathbf{\tilde{f}}|\left\{\tilde{\mathbf{x}} \right\},\mathbf{y},X\right]\\
	&=\sum_{p=1}^{P}k\left(\tilde{\mathbf{x}},\mathbf{x}^{p}\right)\mathbf{\omega}_{p},
	\label{GPR_SUM}\end{align}
	where $\boldsymbol{\omega}=\left( \omega_{1},\dots,\omega_{1}\right)^{\top} $ is a vector of weights determined by 
	\begin{equation}
	\boldsymbol{\omega}=\left[K\left(X,X\right)+\sigma_{P}^{2}I_{P}\right]^{-1}\mathbf{y}.
	\end{equation} 
	
	The computation in \eqref{eq:prediction} requires the knowledge of
	the covariance function $K$ and of the noise variance $\sigma_{P}^{2}$.
	A commonly used covariance function is the Matern 3/2  kernel $k_{Ma}:\mathbb{R}^{d}\times\mathbb{R}^{d}\rightarrow\mathbb{R}$, which is given by
	\begin{equation}\label{cov_func}
	k_{Ma}\left(\mathbf{x},\mathbf{x}'\right)=\sigma_{f}^{2}\left(1+\frac{\sqrt{3} \left\Vert \mathbf{x}-\mathbf{x}'\right\Vert _{2} }{\sigma_{l}} \right)\exp\left(-\frac{\sqrt{3} \left\Vert \mathbf{x}-\mathbf{x}'\right\Vert _{2}}{\sigma_{l}} \right)\mathrm{\ \ for\ } \mathbf{x},\mathbf{x}'\in\mathbb{R}^{d},
	\end{equation}
	where $\sigma_{f}^{2}$ is called the signal variance and $\sigma_{l}$
	is called the length-scale. Another possible choice is the Squared Exponential kernel $k_{SE}:\mathbb{R}^{d}\times\mathbb{R}^{d}\rightarrow\mathbb{R}$,
	which is given by
	\begin{equation}
	k_{SE}\left(\mathbf{x},\mathbf{x}'\right)=\sigma_{f}^{2}\exp\left(-\frac{
	\left\Vert \mathbf{x}-\mathbf{x}'\right\Vert _{2}^{2} }{ 2\sigma_{l}^{2} }\right)  \mathrm{\ \ for\ }  \mathbf{x},\mathbf{x}'\in\mathbb{R}^{d}.\label{SEk}
	\end{equation}
	In general, the choice of kernel function is performed by using a log-likelihood criterion.  	 The parameters
	$\sigma_{f}^{2}$, $\sigma_{l}$ of the kernel function and
	$\sigma_{P}^{2}$ of the noise are called hyperparameters and need
	to be estimated. A common approach is to consider the maximum likelihood
	estimates which can be obtained by maximizing the log-likelihood function
	of the training data, that is by maximizing the following function:
	\begin{equation}
	-\frac{1}{2}\log\left(\det\left(K\left(X,X\right)+\sigma_{P}^{2}I_{P}\right)\right)-\frac{1}{2}\mathbf{y}^{\top}\left[K\left(X,X\right)+\sigma_{P}^{2}I_{P}\right]^{-1}\mathbf{y}.
	\end{equation}

	The development of the GPR model can be divided in the training step
	and the evaluation step (also called testing step). The training step
	only requires the knowledge of the training set $\mathcal{D}$ and
	it consists in estimating  the hyperparameters
	and computing the vector of weights $\boldsymbol{\omega}$.
	The evaluation step can be computed only after the training step has
	been accomplished and it consists in obtaining the predictions via
	the computation of $K\left(\tilde{X},X\right) \boldsymbol{\omega}$.
	We stress out that the training step is independent of the test set
	$\tilde{X}$. Thus one can store the values computed during the training
	step and perform the evaluation step many times with a small computational
	cost, which is $\mathcal{O}\left(P\cdot M\right)$.
	\begin{remark}\label{rm1}
 We observe that the computation time depends only marginally on the size $d$ of the space where the points lie, as the value of $d$ only impacts in the time taken to calculate distances between the points which appears in the covariance matrix K, that is $\left\Vert \mathbf{x}-\mathbf{x}'\right\Vert _{2}$.
\end{remark}
	
\subsection{Machine Learning Exact Integration for European options}
In order to improve the GPR-MC approach, we employ the European option price as a control variate. 
	Here, we propose to compute such a price by means of the semi-analytical formula introduced by Gouden\`ege et al. \cite{goudenege2019machine}, that we term GPR-EI formula. This computation is based on two steps. First of all, the payoff function is approximated  by means of  GPR. Then, the European price is computed as the discounted expected value of the final cash flow, that is a multidimensional integral of the payoff function with respect to the log-underlying process density. Such an integral can be computed by means of a closed formula when replacing the true payoff function with its GPR approximation. 

Let us consider a set $Z=\left\{ \mathbf{z}^{q},q=1,\dots,Q\right\} $ consisting of $Q$ points in $\mathbb{R}^{d}$
 quasi-randomly distributed
according to the law of the vector $\left(\sigma_{1}W_{T}^{1},\dots,\sigma_{d}W_{T}^{d}\right)^{\top}$.
In particular, we define 
\begin{equation}
\mathbf{z}_{i}^{q}=\sqrt{T}\sigma_{i}\Sigma_{i}\mathbf{h}^{q},
\end{equation}
where $\Sigma_{i}$ is i-th row of the matrix $\Sigma$ and $\mathbf{h}^{q}$
is the q-th point of the Halton sequence in $\mathbb{R}^{d}$ (other
low-discrepancy sequence can be considered, such as Solob's or Faure's
ones).
Let $u:Z\rightarrow\mathbb{R}$ be the function defined by
\begin{equation}
u\left(\mathbf{z}\right):=\Psi\left(\mathbf{S}_{0}\exp\left(\left(r-\boldsymbol{\eta}-\frac{1}{2}\boldsymbol{\sigma}^{2}\right)T+\mathbf{z}\right)\right).\label{eq:u_def}
\end{equation}
In a nutshell, the main idea is to approximate the function $u$ by
training the GPR method on the  set $Z$. In particular, we employ
the Squared Exponential kernel defined in \eqref{SEk}. Equation \eqref{GPR_SUM} allows one to approximate the function $u\left(\cdot\right)$ by
\begin{equation}
u^{GPR}\left(\mathbf{z}\right)=\sum_{q=1}^{Q}k_{SE}\left(\mathbf{z}^{q},\mathbf{z}\right)\mathbf{\omega}_{q},
\end{equation}
where $\omega_{1},\dots,\omega_{P}$ are weights. The continuation value can be computed
by integrating the function $u^{GPR}$ against a $d$-dimensional
probability density. The use of the Squared Exponential kernel  allows one to easily perform such a calculation  by means of a closed formula. 
Specifically, the GPR-EI method relies on the following Proposition.
\begin{prop}\label{Prop1}
 Let us consider an European option with payoff function $\Psi$, inception $t=0$, maturity $T$, and multidimensional underlying following the dynamics in \eqref{sde} with spot price $\mathbf{S}_0$.	The  price of such an  option at $t=0$ can be approximated	by
	\begin{equation}\label{vEU1}
	v^{EU}=e^{-rT}E_{0,\mathbf{S}_{0}}\left[\Psi\left(\mathbf{S}_{T}\right)\right]\approx e^{-rT}\sum_{q=1}^{Q}\omega_{q}\sigma_{f}^{2}\sigma_{l}^{d}\frac{e^{-\frac{1}{2}\left(\mathbf{z}^{q}\right)^{\top}\left(T\cdot\Pi+\sigma_{l}^{2}I_{d}\right)^{-1}\left(\mathbf{z}^{q}\right)}}{\sqrt{\det\left(T\cdot\Pi+\sigma_{l}^{2}I_{d}\right)}}
	\end{equation}
	where $\sigma_{f}$, $\sigma_{l}$, and $\omega_{1},\dots,\omega_{Q}$
	are certain constants determined by the GPR approximation of the function
	$\mathbf{z}\mapsto u\left(\mathbf{z}\right)$ considering $Z$ as
	the predictor set, and $\Pi=\left(\Pi_{i,j}\right)$ is the $d\times d$
	covariance matrix of the vector $\left(\sigma_{1}W_{T}^{1},\dots,\sigma_{d}W_{T}^{d}\right)^{\top}$, that is $\Pi_{i,j}=\rho_{i,j}\sigma_{i}\sigma_{j} $.
\end{prop} The proof of this Proposition is very similar to the one reported
in \cite{goudenege2019machine}.

\medskip 
Despite the GPR-EI formula \eqref{vEU1} is adapted to compute the option price supposing the spot price to be $\mathbf{S}_0$ and the time to maturity to be $T$, it works quite well also for   spots close to $\mathbf{S}_0$ and time to maturity smaller than $T$. The following Proposition states how to do that.

\begin{prop}\label{Prop2}
	Let us consider and European option with payoff function $\Psi$, inception $0<\tilde{t}<T$, maturity $T$, and multidimensional underlying following the dynamics in \eqref{sde}. Let  $\tilde{\mathbf{S}}$ be the vector of the spot prices at time $\tilde{t}$ and define $\tilde{\mathbf{z}}\in \mathbb{R}^d$ such that 
	\begin{equation}
\tilde{\mathbf{S}}=\mathbf{S}_{0}\exp\left(\left(r-\boldsymbol{\eta}-\frac{1}{2}\boldsymbol{\sigma}^{2}\right)\tilde{t}+\tilde{\mathbf{z}}\right) .
	\end{equation} 
	The  price of such an  option at $\tilde{t}$ can be approximated	by
	\begin{equation}\label{vEU2}
	v^{EU}=e^{-r(T-\tilde{t})}E_{\tilde{t},\tilde{\mathbf{S}}}\left[\Psi\left(\mathbf{S}_{T}\right)\right]\approx e^{-r(T-\tilde{t})}\sum_{q=1}^{Q}\omega_{q}\sigma_{f}^{2}\sigma_{l}^{d}\frac{e^{-\frac{1}{2}\left(\mathbf{z}^{q}-\tilde{\mathbf{z}}\right)^{\top}\left((T-\tilde{t})\cdot\Pi+\sigma_{l}^{2}I_{d}\right)^{-1}\left(\mathbf{z}^{q}-\tilde{\mathbf{z}}\right)}}{\sqrt{\det\left((T-\tilde{t})\cdot\Pi+\sigma_{l}^{2}I_{d}\right)}}
	\end{equation}
	where $\sigma_{f}$, $\sigma_{l}$, and $\omega_{1},\dots,\omega_{Q}$
	 and $\Pi=\left(\Pi_{i,j}\right)$ are defined according to Proposition \ref{Prop1}.
\end{prop}
The proof of   Proposition \ref{Prop2} derives directly from Proposition \ref{Prop1} by considering $\tilde{\mathbf{S}}$ in place of $ \mathbf{S}_0$. The hyperparameters   $\sigma_{f}$, $\sigma_{l}$, and the weights   $\omega_{1},\dots,\omega_{Q}$
need to be computed only once and then we can use formulas \eqref{vEU1} and \eqref{vEU2} to compute the European prices. The resolution of the linear systems within the exponential factors and the computation of the matrix determinant in \eqref{vEU1} and \eqref{vEU2} can be done quite fast by computing the Cholesky decomposition of the matrices $(T-\tilde{t})\cdot\Pi+\sigma_{l}^{2}I_{d}$ for each of the few possible values of $t$, that is $t=0,t_1,\dots,t_{N-1}$.
 For this reason, turns out to be faster than repeated Monte Carlo simulations to compute the many European prices to be used as control variate.

\subsection{Machine Learning Control Variate algorithm for  American options}
\subsubsection{\label{Sec31}The GPR Monte Carlo Method}
Let us introduce the GPR Monte Carlo approach.
We approximate the price of an American option with the price of a Bermudan option on the same basket.  Specifically, let $N$ be the number of time steps and $\Delta t=T/N$ the time increment. The discrete exercise dates are  $t_n=n\,\Delta t$, as
$n=1,\ldots,N$.
If $\mathbf{x}$ represents the vector of the underlying prices at the exercise date $t_n$, then the price of the Bermudan option $v^{BM}$ is given by
\begin{equation}
v^{BM}\left(t_{n},\mathbf{x}\right)=\max\left(\Psi\left(\mathbf{x}\right),E_{t_{n},\mathbf{x}}\left[e^{-r\Delta t}v\left(t_{n+1},\mathbf{S}_{t_{n+1}}\right)\right]\right). \label{eq:update}
\end{equation}
First of all, by knowing the function $v^{BM}\left(t_{n+1},\cdot\right)$, one can compute   $v^{BM}\left(t_{n},\cdot\right)$ by approximation 
of the expectation in  \eqref{eq:update}.
In order to do that, we consider a set $X^{n}$ of $P$ points whose coordinates represent certain possible values for the underlyings at time $t_n$: 
\begin{equation} 
X^{n}=\left\{ \mathbf{x}^{n,p}=\left(x_{1}^{n,p},\dots,x_{d}^{n,p}\right),p=1,\dots,P\right\} \subset\mathbb{R}^{d}.
\end{equation}
 Suppose now we want to compute  $v^{BM}\left(t_{n},\cdot \right)$ but only for $\mathbf{x}^{n,p}\in X^{n}$.
This goal can be achieved by means of a one step Monte Carlo simulation. In particular, for each $\mathbf{x}^{n,p}\in X^n$, we simulate a set of points $\tilde{X}^{n}_{p}$   
\begin{equation} 
\tilde{X}^{n}_{p}=\left\{ \mathbf{\tilde{x}}^{n,p,m}=\left(\tilde{x}_{1}^{n,p,m},\dots,\tilde{x}_{d}^{n,p,m}\right),m=1,\dots,M\right\} \subset\mathbb{R}^{d}
\end{equation}
of $M$  possible values for $\mathbf{S}_{t_{n+1}}$ according to the law of $\mathbf{S}_{t_{n+1}}\left|\ensuremath{\mathbf{S}_{t_{n}}=\mathbf{x}}\right.$. In particular, for $i=1,\dots,d$, $n=1,\dots,N$, $p=1,\dots,P$, $m=1,\dots,M$, we define
\begin{equation}\label{xtilde}
 \tilde{x} ^{n,p,m}_{i} =x^{n,p}_{i} e^{\left(r-\eta_{i} - \frac{1}{2} {\sigma}^{2}_{i}\right) \Delta t+\sqrt{\Delta t}\sigma_{i} \Sigma_i  \mathbf{G}^{n,p,m}},
\end{equation}
where $\mathbf{G}^{n,p,m}\sim \mathcal{N}\left(0,I_d\right)$  is a standard Gaussian random vector and $\Sigma_i$ is the $i$-th row of the matrix $\Sigma$, just as in  \eqref{sde_cho}.
Then, the option price can be approximated for each $\mathbf{x}^{n,p}\in X^n$  by
\begin{equation}
\hat{v}^{BM}\left(t_{n},\mathbf{x}^{n,p}\right)=\max\left(\Psi\left(\mathbf{x}^{n,p}\right),\frac{e^{-r\Delta t}}{M}\sum_{m=1}^{M}v^{BM}\left(t_{n+1},\mathbf{\tilde{x}}^{n,p,m}\right)\right), \label{eq:update2}
\end{equation}
if the quantities $v^{BM}\left(t_{n+1},\mathbf{\tilde{x}}^{n,p,m}\right)$ are known for all of these simulated points $\mathbf{\tilde{x}}^{n,p,m}$. If we proceed backward, the function ${v}^{BM}\left(t,\cdot \right)$ is known for $t=T$ since it is equal to the payoff function $\Psi\left(\cdot\right) $ and thanks to \eqref{eq:update2} it is known, through an approximation, also for $t=t_{N-1}$ and $\mathbf{x}^{N-1,p}\in X^{N-1}$. In order to assess ${v}^{BM}\left(t_{N-2},\mathbf{x}^{N-2,p} \right)$ for all $\mathbf{x}^{N-2,p}\in X^{N-2}$, and thus going on up to $t = 0$, it is necessary to evaluate the function ${v}^{BM}\left(t_{N-2},\cdot \right)$ for all the points  in $\tilde{X}^{N-2}=\bigcup_{p=1}^{P}\tilde{X}^{N-2,p}$. This cannot be done directly since we know  $\hat{v}^{BM}\left(t_{N-1},\cdot \right)$ only for the points in $ X^{N-1}$ and not for all those in $\tilde{X}^{N-2}$.
To overcome this issue, we compute the approximation of the function $\hat{v}^{BM}\left(t_{N-1},\cdot \right)$ by means of the GPR technique. In particular the set $X^{N-1}$ serves as the predictor set and $\left\lbrace \hat{v}^{BM}\left( t_{N-1}, \mathbf{x}^{N-1,p}\right) ,p=1,\dots,P \right\rbrace $ as the response set.

 More generally, let ${v}^{BM,GPR}_{n}\left(\cdot \right)$ be the GPR approximation of $\hat{v}^{BM}\left(t_{n},\cdot \right)$ trained by considering $X^{n}$ as the predictor set and $\left\lbrace \hat{v}^{BM}\left( t_{n}, \mathbf{x}^{n,p}\right) ,p=1,\dots,P \right\rbrace $ as the response set, where $\hat{v}^{BM}$ is defined as in \eqref{eq:update2}. Then, we can proceed backward by computing 
\begin{equation}\label{back_step}
\hat{v}^{BM}\left(t_{n-1},\mathbf{x}^{n-1,p}\right)=\max\left(\Psi\left(\mathbf{x}^{n-1,p}\right),\frac{e^{-r\Delta t}}{M}\sum_{m=1}^{M}v^{BM,GPR}_{n}\left(\mathbf{\tilde{x}}^{n-1,p,m}\right)\right). 
\end{equation}
and by computing $v^{BM,GPR}_{n-1}$, that is the GPR approximation of $\hat{v}^{BM}\left( t_{n-1},\cdot \right) $. 
Finally, the option price at time $t=0$   is computed through
\begin{equation}\label{P0}
\hat{v}^{BM}\left(0,\mathbf{S}_0\right)=\max\left(\Psi\left(\mathbf{S}_{0}\right),\frac{e^{-r\Delta t}}{M}\sum_{m=1}^{M} \tilde{v}_{1}^{BM,GPR}\left(\mathbf{\tilde{x}}^{0,m}\right)\right)
\end{equation}
where the points $\mathbf{\tilde{x}}^{0,1},\dots, \mathbf{\tilde{x}}^{0,M}$ are random simulations of $\mathbf{S}_{t_1}$ given by 

\begin{equation} 
\tilde{x} ^{0,m}_{i} =S_{0}^{i} e^{\left(r-\eta_{i} - \frac{1}{2} {\sigma}^{2}_{i}\right) \Delta t+\sqrt{\Delta t}\sigma_{i} \Sigma_i  \mathbf{G}^{0,m}},
\end{equation}
where $\mathbf{G}^{0,m}\sim \mathcal{N}\left(0,I_d\right)$  is a standard Gaussian random vector for any $m\in\left\lbrace 1,\dots,M\right\rbrace $.

The choice of the sets $X^{n},n=1\dots,N-1$ is a sensitive question. 
Similarly to what proposed by Ludkovski \cite{ludkovski2018Kriging}, here we use a  deterministic space-filling sequence based on the Halton sequence.
Specifically, let $\mathbf{h}^{p}$ be the $p$-th point of the Halton quasi-random sequence in $\mathbb{R}^d$ and $\Phi^{-1}$ the  inverse cumulative distribution of a standard normal distribution.
We define the points $\mathbf{x}^{n,p}$ as follows: 
\begin{equation}\label{x}
\mathbf{x}^{n,p}_{i} =\mathbf{S}_{0}^{i} e^{\left(r-\eta_{i} - \frac{1}{2} {\sigma}^{2}_{i}\right) t_{n}+\sqrt{t_n}\sigma_{i} \Sigma \Phi^{-1}\left( H^{p}\right)},
\end{equation}
for $i=1,\dots,d$, $n=1,\dots,N-1$, and  $p=1,\dots,P$. 
This choice for the sets $X^{n}$ proves to be the most effective, since the points used to train the GPR algorithm at time $t_n$ are sampled according to the density function of the process $\mathbf{S}_{t_{n}}$.

\subsubsection{The GPR Monte Carlo Control Variate Method}
Let us present the GPR Monte Carlo Control Variate method (GPR-MC-CV), that is our proposed algorithm. 

The control variate technique  is commonly used to reduce the variance of Monte Carlo estimators, but it can also give its contribution in American pricing. Following Bally et al. \cite{bally2005pricing} and  Caramellino and Zanette \cite{caramellino2011monte}, we employ the European price as a control variate for the American price. Let us consider an American and an European option  with the same payoff function $\Psi$ and maturity $T$, and let     $v^{AM},v^{EU}$ denote their prices  respectively. For a fixed time $t$ and underlying stocks $\mathbf{x}$, we define the American-European price gap as:  
\begin{equation}
v\left(t,\mathbf{x}\right)=v^{AM}\left(t,\mathbf{x}\right)-v^{EU}\left(t,\mathbf{x}\right).
\end{equation}
Then 
\begin{equation}
v\left(T,\mathbf{x}\right)=0,
\end{equation}
and it is straightforward to see that
\begin{equation}
v\left(t,\mathbf{x}\right)=\sup_{\tau\in\mathcal{T}_{t,T}}E_{t,\mathbf{x}}\left[e^{-r\left(\tau-t\right)}\hat{\Psi}\left(\tau,\mathbf{S}_{\tau}\right)\right].
\end{equation}
where $\mathcal{T}_{t,T}$ stands for the set of all stopping times taking values in $\left[ t,T\right] $ and $\hat{\Psi}$ is defined by
\begin{equation}\label{Psihat}
\hat{\Psi}\left(t,\mathbf{x}\right)= \Psi\left(\mathbf{x}\right)-v^{EU}\left(t,\mathbf{x}\right).
\end{equation}
We stress out that $\hat{\Psi}\left(T,\mathbf{x}\right)=0$ and the function $\hat{\Psi}$ depends on the time variable also.
Therefore, in order to numerically evaluate $v^{AM}\left(0,\mathbf{S}_{0}\right)$, one can arrange a dynamic programming principle, based on Bermudan approximation, actually equal to the one in Section \ref{Sec31} by replacing $\Psi$ with $\hat{\Psi}$. Once the initial price gap $v \left(0,\mathbf{S}_{0}\right)$ has been calculated, one can retrieve the American price by computing
\begin{equation}
v^{AM}\left(0,\mathbf{S}_{0}\right)=v\left(0,\mathbf{S}_{0}\right)+v^{EU}\left(0,\mathbf{S}_{0}\right).
\end{equation}


 The sketch of the GPR-MC-CV algorithm is presented here.
	
\bigskip
{\small

\framebox[16.4cm][l]{
\begin{minipage}[l]{16.4cm}

\texttt{
\begin{tabbing}
$\quad$ Preprocessing: \= compute   $\mathbf{x}^{n,p}$ and $\mathbf{\tilde{x}}^{n,p,m}$ by using equations \eqref{x} and \eqref{xtilde}, \\
\> $v^{EU}\left( t_n,\mathbf{x}^{n,p}\right) $ and $\hat{\Psi}\left( t_n,\mathbf{x}^{n,p}\right) $ by using \eqref{vEU1}, \eqref{vEU2} and \eqref{Psihat}\\
\medskip
\medskip
$\quad$  Step $N-1$: shaping of $v^{GPR}_{N-1}\left(\cdot\right)$:\\
\> $\hookrightarrow$ \= For $p=1,\dots,P$ compute $\hat{v}\left(t_{N-1},\mathbf{x}^{N-1,p}\right)=\hat{\Psi}\left(\mathbf{x}^{N-1,p}\right)$\\
\> $\hookrightarrow$ \= Define the training set $\mathcal{D}=\left\lbrace \left(\mathbf{x}^p, \hat{v}\left(t_{N-1},\mathbf{x}^{N-1,p}\right)\right),p=1,\dots,P \right\rbrace$ \\
\> $\hookrightarrow$ \= Train GPR  on $\mathcal{D}$ to obtain $v^{GPR}_{N-1}\left(\cdot\right)$\\
\medskip
$\quad$  Step $N-2$: shaping of $v^{GPR}_{N-2}\left(\cdot\right)$:\\
\> $\hookrightarrow$ \= For $p=1,\dots,P$ compute\\
\>\phantom{$\quad$} $\hat{v}\left(t_{N-2},\mathbf{x}^{N-2,p}\right)=\max\left(\hat{\Psi}\left(\mathbf{x}^{N-2,p}\right),\frac{e^{-r\Delta t}}{M}\sum_{m=1}^{M} v^{GPR}_{N-1}\left(\mathbf{\tilde{x}}^{N-2,p,m}\right)\right)$\\
\> $\hookrightarrow$ \= Define the training set $\mathcal{D}=\left\lbrace \left(\mathbf{x}^p, \hat{v}\left(t_{N-2},\mathbf{x}^p\right)\right),p=1,\dots,P \right\rbrace$ \\
\> $\hookrightarrow$ \= Train GPR  on $\mathcal{D}$ to obtain $v^{GPR}_{N-2}\left(\cdot\right)$\\
\end{tabbing}
}
%


\texttt{
$
\quad
\begin{array}{l}
\vdots
\end{array}
$
$\leftarrow$ Steps $n = N-3,\ldots,1$
$
\left[
\begin{array}{l}
\mbox{replace $N-2$ with $n$ and $N-1$ with $n+1$;
}\\
\end{array}
\right]
$
}

\medskip
\medskip

\texttt{
$\quad$
Step $0$: computation of the price:
$$
\hat{v}\left(0,\mathbf{S}_0\right)=\max\left(\Psi\left(\mathbf{S}_{0}\right),\frac{e^{-r\Delta t}}{M}\sum_{m=1}^{M} v^{GPR}_{1} \left( \mathbf{\tilde{x}}^{0,m}\right)\right)
$$
$$
v^{BM}\left(0,\mathbf{S}_0\right)=\hat{v}\left(0,\mathbf{S}_0\right)+v^{EU}\left(0,\mathbf{S}_0\right)
$$
}
\end{minipage}
}

} 
\bigskip
\begin{remark}
	We remark that when using a quasi-Monte Carlo sequence it is important to consider leaping. This technique consists in considering only some uniformly subsampled points of the original sequence, which improves convergence. However,  the leap values,  must be chosen with care.  In fact, many values lead to sequences that do not touch on large sub-hyper-rectangles of the unit hypercube, failing to be a uniform quasi-random point set (see Kocis and Whiten \cite{kocis1997computational}). A common rule for choosing the leap values for the Halton sequence consists in  setting the value to $q-1$, where $q$ is a prime number that has not been used to generate the sequence. 
\end{remark}
\medskip
\begin{remark}
	We observe that the Monte Carlo evaluation of the continuation value    can be easily parallelized since the  summations in \eqref{back_step}, are independent of each other and can be calculated separately. Thus, this feature  allows one to significantly reduce the computational time.
\end{remark}	
\medskip
\begin{remark}
As observed by Ludkovsi \cite{ludkovski2018Kriging}, the main computational cost is	due to the training of the GPR model, which is proportional to the cube of the observation amount. In our case, this training has to be performed one time to compute the European prices with a cost $\mathcal{O}\left(Q^3 \right)$ (with $Q$   the number of points employed in   European price computation), and $N-2$ times within the algorithm to approximate the American-European gap at a give time, thus $\mathcal{O}\left(N\cdot P^3 \right)$ (with $N$  the number of time steps and $P$ the number of points used to train the GPR models at each time step). On the other hand, the cost of the Monte Carlo step depends on both the number of evaluations to be performed and to the number of points employed: the cost for such a step is $\mathcal{O}(N\cdot P \cdot M)$ (with $M$  the number of Monte Carlo simulations employed in estimating the continuation gap value). Finally, we observe that if we compute the European prices by using $M'$  Monte Carlo simulations instead of by using the GPR-EI formula, then the cost  $\mathcal{O}\left(Q^3 \right)$ is replaced by $\mathcal{O}(N\cdot P \cdot M')$.

\end{remark}	
  
\subsubsection{The Control Variate for GPR-Tree and GRP-EI}
Although the control variable technique was initially conceived as a variance reduction techniques for Monte Carlo methods, it can also be a valid support in other contexts.
We investigate the benefits brought by this technique to the GPR-Tree and GPR-EI techniques introduced by Gouden\`ege et al. \cite{goudenege2019machine} for pricing American options in high dimension. In particular, as proposed for the GPR-MC method, we use the European price as a control variate and we employ   GPR-Tree (or   GPR-EI) to compute the American-European price gap.
Let us give a brief introduction of these two numerical approaches. We  refer the interested reader to \cite{goudenege2019machine} for more details.
 
The GPR-Tree method   is similar to the GPR-MC method here proposed. The main difference consists in the use of a tree step  in place of random simulations to compute the continuation value. In particular, for each time step $t_{n}$ and for each point $\mathbf{x}^{p}$, $2^d$ future values are generated according to the tree method proposed by Ekvall \cite{ekvall1996lattice}, in place of Monte Carlo simulations. Such a method is particularly efficient when the dimension $d$ is low (that is, indicatively, it does not exceed 10).
 
The GPR-EI method differs from both the GPR-MC and GPR-Tree methods for three reasons. 
First of all, the predictors employed in the GPR step are related to the logarithms of the underlying value. Then, the continuation value at these points is computed through a  closed formula which comes from an  exact integration. Finally, the GPR-EI method employs the Squared Exponential kernel, which is given by
 	\begin{equation}
 	k\left(\mathbf{x},\mathbf{x}'\right)=\sigma_{f}^{2}\exp\left(-\frac{1}{2 \sigma_{l}^{2}} \left\| \mathbf{x}-\mathbf{x}'\right\| ^{2}_{2}\right),
 	\end{equation}
 	for $\mathbf{x},\mathbf{x}'\in \mathbb{R}^d$, where $d$ is the dimension of the regression problem.

	\section{Numerical Results}

In this Section we report some numerical results in order to investigate
the effectiveness of the proposed Machine Learning algorithm for pricing
American options in the multi-dimensional Black-Scholes model. 

First of all, we compare the  GPR-MC and GPR-MC-CV methods considering Geometric and Arithmetic basket put options and then we focus on a Call on the Maximum option. Moreover, we study the benefits of using the control variable also for GPR-Tree and GPR-EI methods. Finally, we investigate the variance of the price estimators about the two methods.
We stress out that the GPR-Tree method is interesting only for low dimension options: when $d$ exceeds $10$, the method still works, but computational times grow exponentially.

\subsection{\label{PUT}Geometric and Arithmetic Basket Put  Options}
In this test we focus on two payoff that depend on the mean of the underlyings. Specifically, we consider the following payoff examples:
\begin{itemize}
\item Geometric basket Put
\[
\Psi(\mathbf{S}_T)=\left(K - \left( \prod_{i=1}^d S_T^i\right) ^{\frac 1 d}\right)_+,
\]
\item  Arithmetic basket Put 
\[
\Psi(\mathbf{S}_T)=\left(K - \frac 1 d \sum_{i=1}^d S_T^i\right)_+.
\]

\end{itemize}
We consider both the GPR-MC  and the GPR-MC-CV method in order to investigates the benefits induced by the control variate technique.
We consider the same parameters as in \cite{goudenege2019machine}: $T=1$, $S_i=100$, $K=100$, $r =0.05$, equal (null) dividend rates $\eta_i=0.0$, equal volatilities $\sigma_i=0.2$, equal correlations
$\rho_{ij}=0.2$ and $N=10$ exercise dates. Moreover, we consider $P=250,500$ or $1000$ points, $M=10^3,10^4$ or $10^5$ Monte Carlo simulations and $Q=10000$ points for the computation of the European prices with the GPR-EI formula.
As opposed to the other input parameters, we vary the dimension $d$, considering $d=2,\,5,\,10,\,20,\,40$ and $100$.
The algorithm has been implemented in MATLAB and computations have been preformed on a server which employs a $2.40$ GHz Intel$^{{\scriptsize \textregistered}
}$ Xenon$^{{\scriptsize \textregistered}
}$ processor (Gold 6148, Skylake) and 20 GB of RAM.

We present now the numerical results for the two payoff examples.
 First of all, let us present the European results, obtained by means of the GPR-EI formula.
Table \ref{tab:EU1} reports the prices, changing the dimension $d$ and the number of employed points $Q$. Moreover, we also report a Benchmark price computed by Monte Carlo simulation considering $10^6$ samples ($95\%$ confidence intervals are $\pm0.01$ for all the benchmark values.). As we can see with only $1000$ points we can obtain accurate results in any considered dimension. 

\begin{table}
	\begin{centering}
		\setlength\tabcolsep{6pt}%
\begin{tabular}{ccccccccccccc}
	\toprule 
	&  & \multicolumn{5}{c}{Geometric Basket Put} &  & \multicolumn{5}{c}{Arithmetic Basket Put}\tabularnewline
	&  & \multicolumn{4}{c}{GPR-EI} & Bm &  & \multicolumn{4}{c}{GPR-EI} & Bm\tabularnewline
	\cmidrule{3-7} \cmidrule{9-13} 
	$d$ & $P$ & $\phantom{1}250$ & $\phantom{1}500$ & $1000$ & $8000$ &  &  & $\phantom{1}250$ & $\phantom{1}500$ & $1000$ & $8000$ & \tabularnewline
	\midrule
	2 &  & $\underset{\left(2\right)}{4.10}$ & $\underset{\left(3\right)}{4.11}$ & $\underset{\left(15\right)}{4.13}$ & $\underset{\left(44\right)}{4.17}$ & $4.18$ &  & $\underset{\left(2\right)}{3.83}$ & $\underset{\left(1\right)}{3.85}$ & $\underset{\left(12\right)}{3.86}$ & $\underset{\left(41\right)}{3.90}$ & $3.92$\tabularnewline
	5 &  & $\underset{\left(2\right)}{2.90}$ & $\underset{\left(1\right)}{2.98}$ & $\underset{\left(3\right)}{3.01}$ & $\underset{\left(24\right)}{3.04}$ & $3.06$ &  & $\underset{\left(1\right)}{2.49}$ & $\underset{\left(1\right)}{2.57}$ & $\underset{\left(3\right)}{2.60}$ & $\underset{\left(26\right)}{2.63}$ & $2.64$\tabularnewline
	10 &  & $\underset{\left(1\right)}{2.48}$ & $\underset{\left(1\right)}{2.45}$ & $\underset{\left(3\right)}{2.52}$ & $\underset{\left(26\right)}{2.59}$ & $2.59$ &  & $\underset{\left(1\right)}{2.01}$ & $\underset{\left(1\right)}{2.03}$ & $\underset{\left(3\right)}{2.08}$ & $\underset{\left(25\right)}{2.13}$ & $2.14$\tabularnewline
	20 &  & $\underset{\left(1\right)}{2.28}$ & $\underset{\left(1\right)}{2.33}$ & $\underset{\left(4\right)}{2.26}$ & $\underset{\left(31\right)}{2.31}$ & $2.33$ &  & $\underset{\left(1\right)}{1.81}$ & $\underset{\left(1\right)}{1.84}$ & $\underset{\left(4\right)}{1.80}$ & $\underset{\left(26\right)}{1.85}$ & $1.86$\tabularnewline
	40 &  & $\underset{\left(1\right)}{2.12}$ & $\underset{\left(1\right)}{2.18}$ & $\underset{\left(5\right)}{2.21}$ & $\underset{\left(44\right)}{2.17}$ & $2.20$ &  & $\underset{\left(1\right)}{1.73}$ & $\underset{\left(1\right)}{1.74}$ & $\underset{\left(4\right)}{1.73}$ & $\underset{\left(37\right)}{1.71}$ & $1.72$\tabularnewline
	100 &  & $\underset{\left(1\right)}{2.03}$ & $\underset{\left(1\right)}{2.07}$ & $\underset{\left(7\right)}{2.09}$ & $\underset{\left(43\right)}{2.08}$ & $2.11$ &  & $\underset{\left(1\right)}{1.93}$ & $\underset{\left(1\right)}{1.63}$ & $\underset{\left(5\right)}{1.67}$ & $\underset{\left(35\right)}{1.62}$ & $1.63$\tabularnewline
	\bottomrule
\end{tabular}
		\par\end{centering}
	\caption{\label{tab:EU1} European price results for the Geometric and Arithmetic Basket Put option obtained by  using the GPR-EI formula. In the last  column the prices obtained by using a Monte Carlo simulation. The values in brackets
	are the computational times (in seconds).}
\end{table}

\medskip
Let us now focus on the American results. As far as the Geometric basket Put is considered, it is possible to reduce the problem of
pricing in the $d$-dimensional model to a one dimensional American Put option in the
Black-Scholes model with opportune parameters. The price of such a one dimensional American option   can be computed in a easy way, for example by using the CRR algorithm with $1000$ steps (see Cox et al. \cite{cox1979option}).
Therefore in this case we have a reliable  benchmark to test the algorithm. Moreover, when $d$ is smaller than $10$  we can also compute the price by means of a multi-dimensional binomial tree (see Ekvall et al. \cite{ekvall1996lattice}). In particular, the number of steps employed for the binomial tree is equal to $200$ when $d=2$ and to $50$ when $d=5$. For values of $d$ larger than $5$, prices cannot be approximated via such a tree,  because the memory required for the calculations would be too large.
Results are reported in Tables \ref{tab:LMC_GEO} and \ref{tab:GPR_MC_GEO}. 
 We observe that both the two algorithms are very accurate in low dimension, despite we are approximating an American option with a Bermudan one. When larger baskets are considered, say $d\geq40$, the prices obtained with the GPR-MC are less accurate and less stable while changing the number of points $P$ and the number of Monte Carlo simulations $M$.  The computer processing time of the GPR-MC-CV method are a little higher than those of the GPR-MC  because European prices need to be computed. 

We also stress out that the computer processing time increase little with the size of the problem. This is due to the fact that the dimension affects significantly only the computational time of the Monte Carlo step while the GPR step is only minimally distressed (see Remark \ref{rm1}).

Table \ref{tab:GPR_TE} and \ref{tab:GPR_TE_CV} report the  results for the GPR-Tree and GPR-EI methods employing or not the control variate technique.  By comparing the results of the two Tables, we observe that the option prices for $d\leq10$ are very similar: in this case variate control technique is not crucial to improve convergence. As opposed to that, GPR-EI benefits sensitively from control variate technique when high values of $d$ are considered.

 \begin{table}
 	\begin{centering}

 		\begin{tabular}{cccccccccccccccc}
 			\toprule 
 			& $P$ & \multicolumn{3}{c}{$250$} &  & \multicolumn{3}{c}{$500$} &  & \multicolumn{3}{c}{$1000$} &  &  & \tabularnewline
 			\cmidrule{3-5} \cmidrule{7-9} \cmidrule{11-13} 
 			$d$ &  & $10^{3}$ & $10^{4}$ & $10^{5}$ &  & $10^{3}$ & $10^{4}$ & $10^{5}$ &  & $10^{3}$ & $10^{4}$ & $10^{5}$ &  & Ekvall & Bm\tabularnewline
 			\midrule
 			2 &  & $\underset{\left(8\right)}{4.54}$ & $\underset{\left(36\right)}{4.57}$ & $\underset{\left(347\right)}{4.58}$ &  & $\underset{\left(19\right)}{4.57}$ & $\underset{\left(140\right)}{4.59}$ & $\underset{\left(1340\right)}{4.57}$ &  & $\underset{\left(72\right)}{4.52}$ & $\underset{\left(605\right)}{4.60}$ & $\underset{\left(5121\right)}{4.57}$ &  & $4.62$ & $4.62$\tabularnewline
 			5 &  & $\underset{\left(8\right)}{3.55}$ & $\underset{\left(42\right)}{3.43}$ & $\underset{\left(498\right)}{3.44}$ &  & $\underset{\left(25\right)}{3.43}$ & $\underset{\left(146\right)}{3.45}$ & $\underset{\left(1378\right)}{3.44}$ &  & $\underset{\left(56\right)}{3.45}$ & $\underset{\left(508\right)}{3.42}$ & $\underset{\left(4761\right)}{3.43}$ &  & $3.44$ & $3.45$\tabularnewline
 			10 &  & $\underset{\left(9\right)}{2.99}$ & $\underset{\left(46\right)}{3.03}$ & $\underset{\left(431\right)}{3.02}$ &  & $\underset{\left(29\right)}{2.96}$ & $\underset{\left(142\right)}{2.98}$ & $\underset{\left(1517\right)}{2.95}$ &  & $\underset{\left(55\right)}{2.98}$ & $\underset{\left(616\right)}{2.95}$ & $\underset{\left(5281\right)}{2.96}$ &  &  & $2.97$\tabularnewline
 			20 &  & $\underset{\left(10\right)}{2.68}$ & $\underset{\left(85\right)}{2.68}$ & $\underset{\left(463\right)}{2.69}$ &  & $\underset{\left(30\right)}{2.75}$ & $\underset{\left(210\right)}{2.70}$ & $\underset{\left(1441\right)}{2.72}$ &  & $\underset{\left(64\right)}{2.72}$ & $\underset{\left(597\right)}{2.69}$ & $\underset{\left(5598\right)}{2.70}$ &  &  & $2.70$\tabularnewline
 			40 &  & $\underset{\left(14\right)}{2.71}$ & $\underset{\left(104\right)}{2.58}$ & $\underset{\left(621\right)}{2.58}$ &  & $\underset{\left(24\right)}{2.60}$ & $\underset{\left(263\right)}{2.61}$ & $\underset{\left(2094\right)}{2.62}$ &  & $\underset{\left(74\right)}{2.51}$ & $\underset{\left(655\right)}{2.55}$ & $\underset{\left(6373\right)}{2.54}$ &  &  & $2.56$\tabularnewline
 			100 &  & $\underset{\left(26\right)}{2.50}$ & $\underset{\left(110\right)}{2.51}$ & $\underset{\left(1822\right)}{2.50}$ &  & $\underset{\left(42\right)}{2.48}$ & $\underset{\left(321\right)}{2.45}$ & $\underset{\left(3817\right)}{2.45}$ &  & $\underset{\left(112\right)}{2.43}$ & $\underset{\left(892\right)}{2.45}$ & $\underset{\left(12410\right)}{2.43}$ &  &  & $2.47$\tabularnewline
 			\bottomrule
 		\end{tabular}
 		\par\end{centering}
 	\caption{\label{tab:LMC_GEO}American price results for a Geometric basket Put option obtained by using the
 		GPR-MC method. In the last column the exact benchmark. The values in brackets 		are the computational times (in seconds).}
 	
 \vspace{7mm} 
 
	\begin{centering}

		\begin{tabular}{cccccccccccccccc}
			\toprule 
			& $P$ & \multicolumn{3}{c}{$250$} &  & \multicolumn{3}{c}{$500$} &  & \multicolumn{3}{c}{$1000$} &  &  & \tabularnewline
			\cmidrule{3-5} \cmidrule{7-9} \cmidrule{11-13} 
			$d$ &  & $10^{3}$ & $10^{4}$ & $10^{5}$ &  & $10^{3}$ & $10^{4}$ & $10^{5}$ &  & $10^{3}$ & $10^{4}$ & $10^{5}$ &  & Ekvall & Bm\tabularnewline
			\midrule
			2 &  & $\underset{\left(19\right)}{4.58}$ & $\underset{\left(45\right)}{4.57}$ & $\underset{\left(382\right)}{4.57}$ &  & $\underset{\left(37\right)}{4.58}$ & $\underset{\left(133\right)}{4.57}$ & $\underset{\left(1445\right)}{4.57}$ &  & $\underset{\left(73\right)}{4.57}$ & $\underset{\left(656\right)}{4.57}$ & $\underset{\left(6420\right)}{4.57}$ &  & $4.62$ & $4.62$\tabularnewline
			5 &  & $\underset{\left(19\right)}{3.41}$ & $\underset{\left(43\right)}{3.41}$ & $\underset{\left(387\right)}{3.41}$ &  & $\underset{\left(33\right)}{3.41}$ & $\underset{\left(129\right)}{3.40}$ & $\underset{\left(1445\right)}{3.41}$ &  & $\underset{\left(79\right)}{3.40}$ & $\underset{\left(649\right)}{3.41}$ & $\underset{\left(5206\right)}{3.41}$ &  & $3.44$ & $3.45$\tabularnewline
			10 &  & $\underset{\left(18\right)}{2.95}$ & $\underset{\left(44\right)}{2.95}$ & $\underset{\left(423\right)}{2.95}$ &  & $\underset{\left(35\right)}{2.94}$ & $\underset{\left(157\right)}{2.95}$ & $\underset{\left(1680\right)}{2.95}$ &  & $\underset{\left(78\right)}{2.94}$ & $\underset{\left(764\right)}{2.94}$ & $\underset{\left(5782\right)}{2.95}$ &  &  & $2.97$\tabularnewline
			20 &  & $\underset{\left(22\right)}{2.68}$ & $\underset{\left(52\right)}{2.69}$ & $\underset{\left(523\right)}{2.68}$ &  & $\underset{\left(34\right)}{2.69}$ & $\underset{\left(174\right)}{2.69}$ & $\underset{\left(1629\right)}{2.69}$ &  & $\underset{\left(86\right)}{2.71}$ & $\underset{\left(687\right)}{2.71}$ & $\underset{\left(5950\right)}{2.71}$ &  &  & $2.70$\tabularnewline
			40 &  & $\underset{\left(28\right)}{2.55}$ & $\underset{\left(71\right)}{2.54}$ & $\underset{\left(752\right)}{2.54}$ &  & $\underset{\left(50\right)}{2.55}$ & $\underset{\left(233\right)}{2.54}$ & $\underset{\left(2028\right)}{2.55}$ &  & $\underset{\left(121\right)}{2.55}$ & $\underset{\left(760\right)}{2.56}$ & $\underset{\left(8616\right)}{2.55}$ &  &  & $2.56$\tabularnewline
			100 &  & $\underset{\left(46\right)}{2.46}$ & $\underset{\left(117\right)}{2.46}$ & $\underset{\left(1398\right)}{2.46}$ &  & $\underset{\left(82\right)}{2.50}$ & $\underset{\left(322\right)}{2.48}$ & $\underset{\left(4211\right)}{2.48}$ &  & $\underset{\left(175\right)}{2.47}$ & $\underset{\left(875\right)}{2.48}$ & $\underset{\left(10171\right)}{2.48}$ &  &  & $2.47$\tabularnewline
			\bottomrule
		\end{tabular}
		\par\end{centering}
	\caption{\label{tab:GPR_MC_GEO}American price results for a Geometric basket Put option obtained by using the
		GPR-MC-CV method. In the last column the exact benchmark. The values in brackets 		are the computational times (in seconds).}
\end{table}

\medskip

\begin{table}
	\begin{centering}
\begin{tabular}{cccccccccccc}
	\toprule 
	&  & \multicolumn{3}{c}{GPR-Tree} &  & \multicolumn{3}{c}{GPR-EI} &  &  & \tabularnewline
	\cmidrule{3-5} \cmidrule{7-9} 
	$d$ & $P$ & $250$ & $500$ & $1000$ &  & $250$ & $500$ & $1000$ &  & Ekvall & Bm\tabularnewline
	\midrule
	2 &  & $\underset{\left(4\right)}{4.61}$ & $\underset{\left(7\right)}{4.61}$ & $\underset{\left(22\right)}{4.61}$ &  & $\underset{\left(4\right)}{4.58}$ & $\underset{\left(9\right)}{4.58}$ & $\underset{\left(26\right)}{4.57}$ &  & $4.62$ & $4.62$\tabularnewline
	5 &  & $\underset{\left(9\right)}{3.44}$ & $\underset{\left(15\right)}{3.43}$ & $\underset{\left(23\right)}{3.44}$ &  & $\underset{\left(4\right)}{3.40}$ & $\underset{\left(14\right)}{3.43}$ & $\underset{\left(27\right)}{3.41}$ &  & $3.44$ & $3.45$\tabularnewline
	10 &  & $\underset{\left(10\right)}{3.00}$ & $\underset{\left(33\right)}{2.96}$ & $\underset{\left(60\right)}{2.93}$ &  & $\underset{\left(4\right)}{2.85}$ & $\underset{\left(9\right)}{2.88}$ & $\underset{\left(30\right)}{2.93}$ &  &  & $2.97$\tabularnewline
	20 &  &  &  &  &  & $\underset{\left(4\right)}{2.63}$ & $\underset{\left(9\right)}{2.73}$ & $\underset{\left(29\right)}{2.63}$ &  &  & $2.70$\tabularnewline
	40 &  &  &  &  &  & $\underset{\left(4\right)}{2.45}$ & $\underset{\left(10\right)}{2.52}$ & $\underset{\left(38\right)}{2.53}$ &  &  & $2.56$\tabularnewline
	100 &  &  &  &  &  & $\underset{\left(5\right)}{2.27}$ & $\underset{\left(15\right)}{2.32}$ & $\underset{\left(45\right)}{2.39}$ &  &  & $2.47$\tabularnewline
	\bottomrule
\end{tabular}
\par\end{centering}
\caption{\label{tab:GPR_TE}American price results for a Geometric basket Put option obtained by using the GPR-Tree and GPR-EI methods (without control variate). In the last column the exact benchmark. The values in brackets 		are the computational times (in seconds). }
 	
\vspace{7mm} 

	\begin{centering}
		\begin{tabular}{cccccccccccc}
			\toprule 
			&  & \multicolumn{3}{c}{GPR-Tree} &  & \multicolumn{3}{c}{GPR-EI} &  &  & \tabularnewline
			\cmidrule{3-5} \cmidrule{7-9} 
			$d$ & $P$ & $250$ & $500$ & $1000$ &  & $250$ & $500$ & $1000$ &  & Ekvall & Bm\tabularnewline
			\midrule
			2 &  & $\underset{\left(16\right)}{4.58}$ & $\underset{\left(33\right)}{4.58}$ & $\underset{\left(64\right)}{4.58}$ &  & $\underset{\left(17\right)}{4.57}$ & $\underset{\left(18\right)}{4.57}$ & $\underset{\left(24\right)}{4.57}$ &  & $4.62$ & $4.62$\tabularnewline
			5 &  & $\underset{\left(15\right)}{3.42}$ & $\underset{\left(19\right)}{3.41}$ & $\underset{\left(35\right)}{3.41}$ &  & $\underset{\left(13\right)}{3.41}$ & $\underset{\left(15\right)}{3.41}$ & $\underset{\left(24\right)}{3.40}$ &  & $3.44$ & $3.45$\tabularnewline
			10 &  & $\underset{\left(18\right)}{2.94}$ & $\underset{\left(31\right)}{2.94}$ & $\underset{\left(72\right)}{2.94}$ &  & $\underset{\left(12\right)}{2.91}$ & $\underset{\left(13\right)}{2.93}$ & $\underset{\left(25\right)}{2.93}$ &  &  & $2.97$\tabularnewline
			20 &  &  &  &  &  & $\underset{\left(13\right)}{2.65}$ & $\underset{\left(19\right)}{2.67}$ & $\underset{\left(41\right)}{2.64}$ &  &  & $2.70$\tabularnewline
			40 &  &  &  &  &  & $\underset{\left(19\right)}{2.54}$ & $\underset{\left(33\right)}{2.57}$ & $\underset{\left(57\right)}{2.54}$ &  &  & $2.56$\tabularnewline
			100 &  &  &  &  &  & $\underset{\left(20\right)}{2.47}$ & $\underset{\left(29\right)}{2.46}$ & $\underset{\left(58\right)}{2.47}$ &  &  & $2.47$\tabularnewline
			\bottomrule
		\end{tabular}
		\par\end{centering}
	\caption{\label{tab:GPR_TE_CV}American price results for a Geometric basket Put option obtained by  using the GPR-Tree and GPR-EI methods (with  control variate technique). In the last column the exact benchmark. The values in brackets 		are the computational times (in seconds). }
\end{table}

 As opposed to the Geometric basket Put option,  we have no method to obtain a fully reliable benchmark when dealing with an Arithmetic basket Put  option. However, for small  values of $d$, a reference  price can be obtained by means of a multidimensional tree method (see Ekvall et al. \cite{ekvall1996lattice}), just as shown for the Geometric case. 
Results are reported in Tables \ref{tab:LMC_ARI}  and \ref{tab:GPR_MC_AR}. The conclusions that we can draw in this case are similar to those for the Geometric case: both the two methods are accurate in low dimension,  while the control variate method is more effective in high dimension.
 
Table \ref{tab:GPR_TE_ARI} and \ref{tab:GPR_TE_CV_ARI} report the  results for the GPR-Tree and GPR-EI methods employing or not the control variate technique.  Just as for the Geometric put option  we observe that the option prices for $d\leq10$ are very similar: in this case variate control technique is not crucial to improve convergence. As opposed to that, control variate technique has an impact on GPR-EI results when high values of $d$ are considered. Anyway, in this case, due to the lack of a benchmark price, it is difficult to draw clear cut conclusions.
\begin{table}
	\begin{centering}
		\begin{tabular}{cccccccccccccccc}
			\toprule 
			& $P$ & \multicolumn{3}{c}{$250$} &  & \multicolumn{3}{c}{$500$} &  & \multicolumn{3}{c}{$1000$} &  &  & \tabularnewline
			\cmidrule{3-5} \cmidrule{7-9} \cmidrule{11-13} 
			$d$ &  & $10^{3}$ & $10^{4}$ & $10^{5}$ &  & $10^{3}$ & $10^{4}$ & $10^{5}$ &  & $10^{3}$ & $10^{4}$ & $10^{5}$ &  & Ekvall & $\phantom{Bm}$\tabularnewline
			\midrule
			2 &  & $\underset{\left(8\right)}{4.34}$ & $\underset{\left(43\right)}{4.37}$ & $\underset{\left(365\right)}{4.38}$ &  & $\underset{\left(21\right)}{4.37}$ & $\underset{\left(145\right)}{4.39}$ & $\underset{\left(1588\right)}{4.37}$ &  & $\underset{\left(62\right)}{4.32}$ & $\underset{\left(767\right)}{4.40}$ & $\underset{\left(5183\right)}{4.37}$ &  & {\small{}$4.42$} & \tabularnewline
			5 &  & $\underset{\left(9\right)}{3.25}$ & $\underset{\left(40\right)}{3.12}$ & $\underset{\left(380\right)}{3.14}$ &  & $\underset{\left(20\right)}{3.12}$ & $\underset{\left(149\right)}{3.14}$ & $\underset{\left(1531\right)}{3.13}$ &  & $\underset{\left(60\right)}{3.14}$ & $\underset{\left(565\right)}{3.11}$ & $\underset{\left(5713\right)}{3.11}$ &  & {\small{}$3.15$} & \tabularnewline
			10 &  & $\underset{\left(11\right)}{2.64}$ & $\underset{\left(40\right)}{2.69}$ & $\underset{\left(419\right)}{2.67}$ &  & $\underset{\left(18\right)}{2.65}$ & $\underset{\left(149\right)}{2.66}$ & $\underset{\left(1551\right)}{2.64}$ &  & $\underset{\left(55\right)}{2.66}$ & $\underset{\left(641\right)}{2.62}$ & $\underset{\left(5149\right)}{2.63}$ &  &  & \tabularnewline
			20 &  & $\underset{\left(17\right)}{2.27}$ & $\underset{\left(68\right)}{2.27}$ & $\underset{\left(631\right)}{2.28}$ &  & $\underset{\left(22\right)}{2.39}$ & $\underset{\left(164\right)}{2.35}$ & $\underset{\left(1626\right)}{2.36}$ &  & $\underset{\left(70\right)}{2.39}$ & $\underset{\left(620\right)}{2.36}$ & $\underset{\left(5817\right)}{2.37}$ &  &  & \tabularnewline
			40 &  & $\underset{\left(16\right)}{2.21}$ & $\underset{\left(94\right)}{2.11}$ & $\underset{\left(780\right)}{2.10}$ &  & $\underset{\left(35\right)}{2.17}$ & $\underset{\left(226\right)}{2.19}$ & $\underset{\left(2165\right)}{2.19}$ &  & $\underset{\left(105\right)}{2.15}$ & $\underset{\left(692\right)}{2.19}$ & $\underset{\left(8739\right)}{2.18}$ &  &  & \tabularnewline
			100 &  & $\underset{\left(20\right)}{1.94}$ & $\underset{\left(110\right)}{1.95}$ & $\underset{\left(1494\right)}{1.94}$ &  & $\underset{\left(34\right)}{1.94}$ & $\underset{\left(306\right)}{1.93}$ & $\underset{\left(3452\right)}{1.92}$ &  & $\underset{\left(91\right)}{1.95}$ & $\underset{\left(884\right)}{1.97}$ & $\underset{\left(9820\right)}{1.95}$ &  &  & \tabularnewline
			\bottomrule
		\end{tabular}
		\par\end{centering}
	\caption{\label{tab:LMC_ARI}American price results for a Arithmetic basket Put option obtained by  using the GPR-MC . In the last column the exact benchmark. The values in brackets 		are the computational times (in seconds).}
 	
\vspace{7mm} 

	\begin{centering}
		\begin{tabular}{cccccccccccccccc}
			\toprule 
			& $P$ & \multicolumn{3}{c}{$250$} &  & \multicolumn{3}{c}{$500$} &  & \multicolumn{3}{c}{$1000$} &  &  & \tabularnewline
			\cmidrule{3-5} \cmidrule{7-9} \cmidrule{11-13} 
			$d$ &  & $10^{3}$ & $10^{4}$ & $10^{5}$ &  & $10^{3}$ & $10^{4}$ & $10^{5}$ &  & $10^{3}$ & $10^{4}$ & $10^{5}$ &  & Ekvall & $\phantom{Bm}$\tabularnewline
			\midrule
			2 &  & $\underset{\left(19\right)}{4.38}$ & $\underset{\left(46\right)}{4.37}$ & $\underset{\left(366\right)}{4.37}$ &  & $\underset{\left(31\right)}{4.38}$ & $\underset{\left(122\right)}{4.37}$ & $\underset{\left(1532\right)}{4.37}$ &  & $\underset{\left(71\right)}{4.38}$ & $\underset{\left(590\right)}{4.37}$ & $\underset{\left(5372\right)}{4.37}$ &  & {\small{}$4.42$} & \tabularnewline
			5 &  & $\underset{\left(15\right)}{3.10}$ & $\underset{\left(40\right)}{3.11}$ & $\underset{\left(378\right)}{3.11}$ &  & $\underset{\left(31\right)}{3.11}$ & $\underset{\left(130\right)}{3.10}$ & $\underset{\left(1514\right)}{3.10}$ &  & $\underset{\left(73\right)}{3.09}$ & $\underset{\left(630\right)}{3.11}$ & $\underset{\left(5338\right)}{3.11}$ &  & {\small{}$3.15$} & \tabularnewline
			10 &  & $\underset{\left(18\right)}{2.62}$ & $\underset{\left(44\right)}{2.62}$ & $\underset{\left(422\right)}{2.62}$ &  & $\underset{\left(33\right)}{2.60}$ & $\underset{\left(129\right)}{2.62}$ & $\underset{\left(1576\right)}{2.61}$ &  & $\underset{\left(78\right)}{2.60}$ & $\underset{\left(668\right)}{2.61}$ & $\underset{\left(5785\right)}{2.61}$ &  &  & \tabularnewline
			20 &  & $\underset{\left(21\right)}{2.33}$ & $\underset{\left(53\right)}{2.35}$ & $\underset{\left(534\right)}{2.34}$ &  & $\underset{\left(36\right)}{2.34}$ & $\underset{\left(182\right)}{2.35}$ & $\underset{\left(1839\right)}{2.35}$ &  & $\underset{\left(86\right)}{2.37}$ & $\underset{\left(656\right)}{2.36}$ & $\underset{\left(5751\right)}{2.36}$ &  &  & \tabularnewline
			40 &  & $\underset{\left(29\right)}{2.19}$ & $\underset{\left(71\right)}{2.19}$ & $\underset{\left(699\right)}{2.19}$ &  & $\underset{\left(50\right)}{2.20}$ & $\underset{\left(255\right)}{2.18}$ & $\underset{\left(2083\right)}{2.19}$ &  & $\underset{\left(113\right)}{2.19}$ & $\underset{\left(870\right)}{2.20}$ & $\underset{\left(7009\right)}{2.19}$ &  &  & \tabularnewline
			100 &  & $\underset{\left(46\right)}{2.09}$ & $\underset{\left(114\right)}{2.09}$ & $\underset{\left(1594\right)}{2.09}$ &  & $\underset{\left(80\right)}{2.12}$ & $\underset{\left(293\right)}{2.09}$ & $\underset{\left(3452\right)}{2.09}$ &  & $\underset{\left(163\right)}{2.07}$ & $\underset{\left(908\right)}{2.09}$ & $\underset{\left(11731\right)}{2.09}$ &  &  & \tabularnewline
			\bottomrule
		\end{tabular}
		\par\end{centering}
 
	\caption{\label{tab:GPR_MC_AR}American price results for a Arithmetic basket Put option obtained by using the GPR-MC-CV method. In the last column the exact benchmark. The values in brackets 		are the computational times (in seconds).}
\end{table}

\begin{table}
	\begin{centering}
		\begin{tabular}{cccccccccccc}
			\toprule 
			&  & \multicolumn{3}{c}{GPR-Tree} &  & \multicolumn{3}{c}{GPR-EI} &  &  & \tabularnewline
			\cmidrule{3-5} \cmidrule{7-9} 
			$d$ & $P$ & $250$ & $500$ & $1000$ &  & $250$ & $500$ & $1000$ &  & Ekvall & $\phantom{Bm}$\tabularnewline
			\midrule
			2 &  & $\underset{\left(5\right)}{4.42}$ & $\underset{\left(9\right)}{4.42}$ & $\underset{\left(25\right)}{4.42}$ &  & $\underset{\left(4\right)}{4.38}$ & $\underset{\left(9\right)}{4.38}$ & $\underset{\left(28\right)}{4.37}$ &  & {\small{}$4.42$} & \tabularnewline
			5 &  & $\underset{\left(5\right)}{3.15}$ & $\underset{\left(9\right)}{3.12}$ & $\underset{\left(24\right)}{3.13}$ &  & $\underset{\left(6\right)}{3.09}$ & $\underset{\left(9\right)}{3.12}$ & $\underset{\left(44\right)}{3.10}$ &  & {\small{}$3.15$} & \tabularnewline
			10 &  & $\underset{\left(10\right)}{2.71}$ & $\underset{\left(21\right)}{2.64}$ & $\underset{\left(70\right)}{2.62}$ &  & $\underset{\left(5\right)}{2.49}$ & $\underset{\left(9\right)}{2.56}$ & $\underset{\left(38\right)}{2.60}$ &  &  & \tabularnewline
			20 &  &  &  &  &  & $\underset{\left(6\right)}{2.26}$ & $\underset{\left(14\right)}{2.31}$ & $\underset{\left(42\right)}{2.28}$ &  &  & \tabularnewline
			40 &  &  &  &  &  & $\underset{\left(4\right)}{2.18}$ & $\underset{\left(10\right)}{2.18}$ & $\underset{\left(31\right)}{2.16}$ &  &  & \tabularnewline
			100 &  &  &  &  &  & $\underset{\left(7\right)}{2.35}$ & $\underset{\left(13\right)}{2.01}$ & $\underset{\left(42\right)}{2.06}$ &  &  & \tabularnewline
			\bottomrule
		\end{tabular}
		\par\end{centering}
	\caption{\label{tab:GPR_TE_ARI}American price results for a Arithmetic basket Put option obtained by using the GPR-Tree and GPR-EI methods (without control variate). In the last column the exact benchmark. The values in brackets 		are the computational times (in seconds). }
 	
\vspace{7mm} 

	\begin{centering}
		\begin{tabular}{cccccccccccc}
			\toprule 
			&  & \multicolumn{3}{c}{GPR-Tree} &  & \multicolumn{3}{c}{GPR-EI} &  &  & \tabularnewline
			\cmidrule{3-5} \cmidrule{7-9} 
			$d$ & $P$ & $250$ & $500$ & $1000$ &  & $250$ & $500$ & $1000$ &  & Ekvall & $\phantom{Bm}$\tabularnewline
			\midrule
			2 &  & $\underset{\left(16\right)}{4.39}$ & $\underset{\left(20\right)}{4.39}$ & $\underset{\left(31\right)}{4.39}$ &  & $\underset{\left(16\right)}{4.37}$ & $\underset{\left(18\right)}{4.37}$ & $\underset{\left(39\right)}{4.37}$ &  & {\small{}$4.42$} & \tabularnewline
			5 &  & $\underset{\left(15\right)}{3.11}$ & $\underset{\left(23\right)}{3.11}$ & $\underset{\left(34\right)}{3.11}$ &  & $\underset{\left(18\right)}{3.10}$ & $\underset{\left(24\right)}{3.10}$ & $\underset{\left(23\right)}{3.10}$ &  & {\small{}$3.15$} & \tabularnewline
			10 &  & $\underset{\left(17\right)}{2.60}$ & $\underset{\left(31\right)}{2.61}$ & $\underset{\left(75\right)}{2.61}$ &  & $\underset{\left(12\right)}{2.57}$ & $\underset{\left(16\right)}{2.59}$ & $\underset{\left(24\right)}{2.59}$ &  &  & \tabularnewline
			20 &  &  &  &  &  & $\underset{\left(14\right)}{2.30}$ & $\underset{\left(19\right)}{2.33}$ & $\underset{\left(31\right)}{2.33}$ &  &  & \tabularnewline
			40 &  &  &  &  &  & $\underset{\left(23\right)}{2.19}$ & $\underset{\left(28\right)}{2.22}$ & $\underset{\left(66\right)}{2.19}$ &  &  & \tabularnewline
			100 &  &  &  &  &  & $\underset{\left(36\right)}{2.16}$ & $\underset{\left(34\right)}{2.13}$ & $\underset{\left(56\right)}{2.13}$ &  &  & \tabularnewline
			\bottomrule
		\end{tabular}
		\par\end{centering}
	\caption{\label{tab:GPR_TE_CV_ARI}American price results for a Arithmetic basket Put option obtained by  using the GPR-Tree and GPR-EI methods (with  control variate technique). In the last column the exact benchmark. The values in brackets 		are the computational times (in seconds). }
\end{table}
 
\subsection{Call on the Maximum option}
Let us consider a Call on the Maximum of $d$-assets American option, whose payoff is given by:
\[
\Psi(\mathbf{S}_T)=\left(\max_{i=1\dots d}  S_T^i-K\right)_+.
\]
 The Call on the Maximum setting is particularly interesting for investigating
scalability of our approaches in the dimension $d$ of the problem. As observed by Ludkovski \cite{ludkovski2018Kriging}, as opposed  to basket Put options, the stopping region of a  Call on the Maximum  consists of several disconnected pieces and this makes the pricing problem particularly challenging.
As done in the previous Section, we consider both the GPR-MC  and the GPR-MC-CV method in order to investigates the benefits induced by the use of this technique.
We consider the same parameters as those employed by Becker et al. \cite{becker2019deep}: $T=3$, $S_i=100$, $K=100$, $r =
0.05$, equal  dividend rates $\eta_i=0.1$, equal volatilities $\sigma_i=0.2$, equal (null) correlations
$\rho_{ij}=0.0$ and $N=9$ exercise dates. Moreover, we consider $P=250,500$ or $1000$ points, $M=10^3,10^4$ or $10^5$ Monte Carlo simulations. As opposed to the other input parameters, we vary the dimension $d$, considering $d=2,\,5,\,10,\,20,\,30,\,50$ and $100$.
In this particular case, because of the long maturity and unbounded payoff,  the GPR-EI formula is not very accurate when considering high dimension and initial points far from the spot $\mathbf{S}_0$, and so we prefer computing the European price by means of Quasi-Monte Carlo simulation with $10^6$ random simulations.

 First of all, let us present the European results, obtained by means of the GPR-EI formula.
Table \ref{tab:EU2} reports the prices, changing the dimension $d$ and the number of employed points $Q$. Moreover, we also report a Benchmark price computed by Monte Carlo simulation considering $10^6$ samples ($95\%$ confidence intervals are $\pm0.01$ for all the benchmark values). 

\begin{table}
	\begin{centering}
		\begin{tabular}{ccccccc}
			\toprule  
			&  & \multicolumn{4}{c}{GPR-EI} & Bm\tabularnewline
			$d$ & $P$ & $\phantom{1}250$ & $\phantom{1}500$ & $1000$ & $8000$ & \tabularnewline
			\midrule
			2 &  & $\underset{\left(1\right)}{10.77}$ & $\underset{\left(2\right)}{10.94}$ & $\underset{\left(12\right)}{10.99}$ & $\underset{\left(89\right)}{11.14}$ & $11.19$\tabularnewline
			5 &  & $\underset{\left(1\right)}{22.36}$ & $\underset{\left(2\right)}{22.68}$ & $\underset{\left(10\right)}{22.82}$ & $\underset{\left(43\right)}{22.99}$ & $23.04$\tabularnewline
			10 &  & $\underset{\left(1\right)}{34.37}$ & $\underset{\left(2\right)}{34.38}$ & $\underset{\left(8\right)}{34.86}$ & $\underset{\left(43\right)}{35.49}$ & $35.59$\tabularnewline
			20 &  & $\underset{\left(1\right)}{48.31}$ & $\underset{\left(1\right)}{49.60}$ & $\underset{\left(7\right)}{48.57}$ & $\underset{\left(44\right)}{49.28}$ & $49.45$\tabularnewline
			30 &  & $\underset{\left(1\right)}{57.47}$ & $\underset{\left(1\right)}{57.46}$ & $\underset{\left(5\right)}{57.05}$ & $\underset{\left(28\right)}{57.62}$ & $57.68$\tabularnewline
			50 &  & $\underset{\left(1\right)}{66.65}$ & $\underset{\left(1\right)}{67.60}$ & $\underset{\left(1\right)}{67.94}$ & $\underset{\left(76\right)}{68.13}$ & $68.03$\tabularnewline
			100 &  & $\underset{\left(1\right)}{80.34}$ & $\underset{\left(1\right)}{81.20}$ & $\underset{\left(5\right)}{81.45}$ & $\underset{\left(34\right)}{82.00}$ & $82.14$\tabularnewline
			\bottomrule
		\end{tabular}
		\par\end{centering}
	\caption{\label{tab:EU2}European price results for a Call on the Maximum option, obtained by  using the GPR-EI formula. In the last  column the prices obtained by using a Monte Carlo simulation. The values in brackets
		are the computational times (in seconds).}
\end{table}
 The aforementioned testing set has also been considered by Becker et al. \cite{becker2019deep} and therefore  we report their results as reference prices.  Furthermore, for small  values of $d$, we can approximate the price obtained by means of a multidimensional tree method. 
Results, which are reported in Tables \ref{tab:LMC_MAX} and \ref{tab:QMC_MC_MAX}, are quite meaningful. Both the two methods perform fine in low dimension, but when large baskets are considered outcomes are strongly different. As far as this particular dataset is considered, the GPR-MC approach gives several null results and others very high, which means that the GPR regression is not able to extrapolate the price surface correctly. In particular, this happens when $d\geq50$. Increasing the number $P$ of points   fixes things when $d=50$ and for $P=1000$ results are likely, although outside the confidence interval proposed by Becker et al. \cite{becker2019deep}. Anyway, when $d=100$ we always obtain null value, showing all the limits of the GPR-MC approach.
As opposed to the GPR-MC, the GPR-MC-CV method performs very well for all the considered dimensions and almost all the values obtained with $P=1000$ and $M=10^5$ are within the confidence intervals proposed by Becker et al. \cite{becker2019deep}.
 
Finally, Tables \ref{tab:GPR_TE_MAX} and \ref{tab:GPR_TE_MAX_CV} report the results for the GPR-Tree and the GPR-EI method obtained by using or not the control variate technique. These two methods seems to be not very effective for the particular Bermudan option considered here. As far as the GPR-EI method is concerned, variates control technique improves the results. Such  a improvement is not evident with respect to the GPR-Tree method.  

\begin{table}
	\begin{centering}
		\begin{tabular}{ccccccccccccccc}
			\toprule 
			& $P$ & \multicolumn{3}{c}{$250$} &  & \multicolumn{3}{c}{$500$} &  & \multicolumn{3}{c}{$1000$} &  & Becker et al.\tabularnewline
			\cmidrule{3-5} \cmidrule{7-9} \cmidrule{11-13} 
			$d$ &  & $10^{3}$ & $10^{4}$ & $10^{5}$ &  & $10^{3}$ & $10^{4}$ & $10^{5}$ &  & $10^{3}$ & $10^{4}$ & $10^{5}$ &  & $95\%\ \text{c.i.}$\tabularnewline
			\midrule
			2 &  & $\underset{\left(10\right)}{14.04}$ & $\underset{\left(32\right)}{13.89}$ & $\underset{\left(404\right)}{13.91}$ &  & $\underset{\left(17\right)}{13.87}$ & $\underset{\left(128\right)}{13.87}$ & $\underset{\left(1156\right)}{13.92}$ &  & $\underset{\left(53\right)}{13.67}$ & $\underset{\left(679\right)}{13.89}$ & $\underset{\left(4479\right)}{13.92}$ &  & $\left[13.88,\,13.91\right]$\tabularnewline
			5 &  & $\underset{\left(8\right)}{26.98}$ & $\underset{\left(36\right)}{26.65}$ & $\underset{\left(362\right)}{26.76}$ &  & $\underset{\left(19\right)}{26.19}$ & $\underset{\left(144\right)}{26.54}$ & $\underset{\left(1633\right)}{26.39}$ &  & $\underset{\left(57\right)}{26.47}$ & $\underset{\left(566\right)}{26.40}$ & $\underset{\left(4837\right)}{26.43}$ &  & $\left[26.14,\,26.17\right]$\tabularnewline
			10 &  & $\underset{\left(9\right)}{38.84}$ & $\underset{\left(42\right)}{38.96}$ & $\underset{\left(430\right)}{38.86}$ &  & $\underset{\left(18\right)}{39.09}$ & $\underset{\left(137\right)}{39.31}$ & $\underset{\left(1363\right)}{39.37}$ &  & $\underset{\left(54\right)}{38.42}$ & $\underset{\left(503\right)}{38.62}$ & $\underset{\left(4304\right)}{38.59}$ &  & $\left[38.30,\,38.37\right]$\tabularnewline
			20 &  & $\underset{\left(10\right)}{60.16}$ & $\underset{\left(72\right)}{59.79}$ & $\underset{\left(378\right)}{59.87}$ &  & $\underset{\left(26\right)}{59.61}$ & $\underset{\left(141\right)}{59.66}$ & $\underset{\left(1431\right)}{59.61}$ &  & $\underset{\left(58\right)}{58.23}$ & $\underset{\left(543\right)}{58.20}$ & $\underset{\left(4835\right)}{58.21}$ &  & $\left[51.55,\,51.80\right]$\tabularnewline
			30 &  & $\underset{\left(15\right)}{73.97}$ & $\underset{\left(73\right)}{73.73}$ & $\underset{\left(439\right)}{73.75}$ &  & $\underset{\left(23\right)}{80.00}$ & $\underset{\left(185\right)}{79.60}$ & $\underset{\left(1789\right)}{79.69}$ &  & $\underset{\left(65\right)}{73.40}$ & $\underset{\left(550\right)}{73.15}$ & $\underset{\left(5192\right)}{73.22}$ &  & $\left[59.48,\,59.87\right]$\tabularnewline
			50 &  & $\underset{\left(15\right)}{93.27}$ & $\underset{\left(84\right)}{93.43}$ & $\underset{\left(1058\right)}{93.26}$ &  & $\underset{\left(27\right)}{21.12}$ & $\underset{\left(234\right)}{21.31}$ & $\underset{\left(2178\right)}{21.30}$ &  & $\underset{\left(69\right)}{113.61}$ & $\underset{\left(604\right)}{113.24}$ & $\underset{\left(6247\right)}{113.18}$ &  & $\left[69.56,\,69.95\right]$\tabularnewline
			100 &  & $\underset{\left(19\right)}{0.01}$ & $\underset{\left(85\right)}{0.01}$ & $\underset{\left(1318\right)}{0.01}$ &  & $\underset{\left(32\right)}{88.44}$ & $\underset{\left(284\right)}{88.05}$ & $\underset{\left(3105\right)}{88.14}$ &  & $\underset{\left(88\right)}{138.42}$ & $\underset{\left(747\right)}{138.18}$ & $\underset{\left(8602\right)}{138.25}$ &  & $\left[83.36,\,83.86\right]$\tabularnewline
			\bottomrule
		\end{tabular}
		\par\end{centering}
	\caption{\label{tab:LMC_MAX}American price results for a Call on the Maximum option obtained by using the GPR-MC . In the last column the exact
		benchmark. The values in brackets are the computational times (in
		seconds). In the last column the confidence intervals reported in \cite{becker2019deep}. }
 	
\vspace{7mm} 

	\begin{centering}
		\begin{tabular}{ccccccccccccccc}
			\toprule 
			& $P$ & \multicolumn{3}{c}{$250$} &  & \multicolumn{3}{c}{$500$} &  & \multicolumn{3}{c}{$1000$} &  & Becker et al.\tabularnewline
			\cmidrule{3-5} \cmidrule{7-9} \cmidrule{11-13} 
			$d$ & $M$ & $10^{3}$ & $10^{4}$ & $10^{5}$ &  & $10^{3}$ & $10^{4}$ & $10^{5}$ &  & $10^{3}$ & $10^{4}$ & $10^{5}$ &  & $95\%\ \text{CI}$\tabularnewline
			\midrule
			2 &  & $ \underset{\left(8\right)}{13.93}$ & $\underset{\left(57\right)}{13.89}$ & $\underset{\left(349\right)}{13.90}$ &  & $\underset{\left(25\right)}{13.89}$ & $\underset{\left(152\right)}{13.88}$ & $\underset{\left(1449\right)}{13.91}$ &  & $\underset{\left(64\right)}{13.86}$ & $\underset{\left(495\right)}{13.90}$ & $\underset{\left(4979\right)}{13.90}$ &  & $\left[13.88,\,13.91\right]$\tabularnewline
			5 &  & $\underset{\left(9\right)}{26.16}$ & $\underset{\left(58\right)}{26.13}$ & $\underset{\left(369\right)}{26.14}$ &  & $\underset{\left(22\right)}{26.07}$ & $\underset{\left(151\right)}{26.12}$ & $\underset{\left(1437\right)}{26.11}$ &  & $\underset{\left(54\right)}{26.13}$ & $\underset{\left(517\right)}{26.13}$ & $\underset{\left(5423\right)}{26.12}$ &  & $\left[26.14,\,26.17\right]$\tabularnewline
			10 &  & $\underset{\left(9\right)}{38.07}$ & $\underset{\left(43\right)}{38.10}$ & $\underset{\left(446\right)}{38.11}$ &  & $\underset{\left(29\right)}{38.20}$ & $\underset{\left(154\right)}{38.21}$ & $\underset{\left(1395\right)}{38.21}$ &  & $\underset{\left(72\right)}{38.26}$ & $\underset{\left(503\right)}{38.29}$ & $\underset{\left(5215\right)}{38.29}$ &  & $\left[38.30,\,38.37\right]$\tabularnewline
			20 &  & $\underset{\left(11\right)}{51.27}$ & $\underset{\left(77\right)}{51.31}$ & $\underset{\left(551\right)}{51.33}$ &  & $\underset{\left(26\right)}{51.60}$ & $\underset{\left(167\right)}{51.64}$ & $\underset{\left(1671\right)}{51.65}$ &  & $\underset{\left(73\right)}{51.56}$ & $\underset{\left(636\right)}{51.58}$ & $\underset{\left(8046\right)}{51.60}$ &  & $\left[51.55,\,51.80\right]$\tabularnewline
			30 &  & $\underset{\left(12\right)}{59.25}$ & $\underset{\left(90\right)}{59.25}$ & $\underset{\left(709\right)}{59.25}$ &  & $\underset{\left(26\right)}{59.33}$ & $\underset{\left(205\right)}{59.32}$ & $\underset{\left(1970\right)}{59.32}$ &  & $\underset{\left(82\right)}{59.54}$ & $\underset{\left(656\right)}{59.54}$ & $\underset{\left(6436\right)}{59.54}$ &  & $\left[59.48,\,59.87\right]$\tabularnewline
			50 &  & $\underset{\left(13\right)}{70.29}$ & $\underset{\left(104\right)}{70.25}$ & $\underset{\left(992\right)}{70.23}$ &  & $\underset{\left(30\right)}{69.61}$ & $\underset{\left(286\right)}{69.57}$ & $\underset{\left(2511\right)}{69.55}$ &  & $\underset{\left(69\right)}{69.73}$ & $\underset{\left(663\right)}{69.69}$ & $\underset{\left(7733\right)}{69.67}$ &  & $\left[69.56,\,69.95\right]$\tabularnewline
			100 &  & $\underset{\left(17\right)}{82.16}$ & $\underset{\left(122\right)}{82.18}$ & $\underset{\left(1621\right)}{82.16}$ &  & $\underset{\left(38\right)}{83.86}$ & $\underset{\left(350\right)}{83.86}$ & $\underset{\left(4196\right)}{83.84}$ &  & $\underset{\left(117\right)}{83.33}$ & $\underset{\left(925\right)}{83.34}$ & $\underset{\left(10506\right)}{83.31}$ &  & $\left[83.36,\,83.86\right]$\tabularnewline
			\bottomrule
		\end{tabular}
		\par\end{centering}
	\caption{\label{tab:QMC_MC_MAX}American price results for  a Call on the Maximum   option obtained by  using the GPR-MC-CV method. In the last column
		the exact benchmark. The values in brackets are the computational
		times (in seconds).  In the last column the confidence intervals reported in \cite{becker2019deep}.}
\end{table}
\begin{table}
	\begin{centering}
		\begin{tabular}{ccccccccccc}
			\toprule 
			&  & \multicolumn{3}{c}{GPR-Tree} &  & \multicolumn{3}{c}{GPR-EI} &  & Becker \tabularnewline
			\cmidrule{3-5} \cmidrule{7-9} 
			$d$ & $P$ & $250$ & $500$ & $1000$ &  & $1000$ & $2000$ & $4000$ &  & et al.\tabularnewline
			\midrule
			2 &  & $\underset{\left(4\right)}{13.83}$ & $\underset{\left(8\right)}{13.83}$ & $\underset{\left(18\right)}{13.85}$ &  & $\underset{\left(19\right)}{13.50}$ & $\underset{\left(49\right)}{13.51}$ & $\underset{\left(53\right)}{13.51}$ &  & $\left[13.88,\,13.91\right]$\tabularnewline
			5 &  & $\underset{\left(5\right)}{25.95}$ & $\underset{\left(9\right)}{25.82}$ & $\underset{\left(22\right)}{25.78}$ &  & $\underset{\left(20\right)}{25.23}$ & $\underset{\left(70\right)}{25.33}$ & $\underset{\left(60\right)}{25.39}$ &  & $\left[26.14,\,26.17\right]$\tabularnewline
			10 &  & $\underset{\left(7\right)}{37.76}$ & $\underset{\left(18\right)}{37.79}$ & $\underset{\left(47\right)}{37.64}$ &  & $\underset{\left(21\right)}{35.90}$ & $\underset{\left(67\right)}{36.69}$ & $\underset{\left(75\right)}{37.09}$ &  & $\left[38.30,\,38.37\right]$\tabularnewline
			20 &  &  &  &  &  & $\underset{\left(23\right)}{46.67}$ & $\underset{\left(73\right)}{49.31}$ & $\underset{\left(100\right)}{49.74}$ &  & $\left[51.55,\,51.80\right]$\tabularnewline
			30 &  &  &  &  &  & $\underset{\left(29\right)}{53.66}$ & $\underset{\left(94\right)}{54.00}$ & $\underset{\left(111\right)}{59.14}$ &  & $\left[59.48,\,59.87\right]$\tabularnewline
			50 &  &  &  &  &  & $\underset{\left(30\right)}{62.17}$ & $\underset{\left(86\right)}{25.84}$ & $\underset{\left(131\right)}{71.86}$ &  & $\left[69.56,\,69.95\right]$\tabularnewline
			100 &  &  &  &  &  & $\underset{\left(32\right)}{70.36}$ & $\underset{\left(145\right)}{74.84}$ & $\underset{\left(262\right)}{51.74}$ &  & $\left[83.36,\,83.86\right]$\tabularnewline
			\bottomrule
		\end{tabular}
		\par\end{centering}
	\caption{\label{tab:GPR_TE_MAX}American price results for a Call on the Maximum option by using the
		GPR-Tree and GPR-EI methods (without control variate). In the last
		column the exact benchmark. The values in brackets are the computational
		times (in seconds).}
 	
\vspace{7mm} 

	\begin{centering}
	\begin{tabular}{ccccccccccc}
		\toprule 
		&  & \multicolumn{3}{c}{GPR-Tree} &  & \multicolumn{3}{c}{GPR-EI} &  & Becker \tabularnewline
		\cmidrule{3-5} \cmidrule{7-9} 
		$d$ & $P$ & $250$ & $500$ & $1000$ &  & $1000$ & $2000$ & $4000$ &  & et al.\tabularnewline
		\midrule
		2 &  & $\underset{\left(20\right)}{13.79}$ & $\underset{\left(48\right)}{13.78}$ & $\underset{\left(77\right)}{13.79}$ &  & $\underset{\left(22\right)}{13.89}$ & $\underset{\left(76\right)}{13.89}$ & $\underset{\left(181\right)}{13.90}$ &  & $\left[13.88,\,13.91\right]$\tabularnewline
		5 &  & $\underset{\left(17\right)}{25.93}$ & $\underset{\left(21\right)}{25.90}$ & $\underset{\left(37\right)}{25.88}$ &  & $\underset{\left(35\right)}{26.04}$ & $\underset{\left(168\right)}{26.08}$ & $\underset{\left(219\right)}{26.12}$ &  & $\left[26.14,\,26.17\right]$\tabularnewline
		10 &  & $\underset{\left(20\right)}{38.73}$ & $\underset{\left(34\right)}{38.73}$ & $\underset{\left(71\right)}{38.62}$ &  & $\underset{\left(49\right)}{38.18}$ & $\underset{\left(239\right)}{38.19}$ & $\underset{\left(297\right)}{38.26}$ &  & $\left[38.30,\,38.37\right]$\tabularnewline
		20 &  &  &  &  &  & $\underset{\left(171\right)}{52.41}$ & $\underset{\left(124\right)}{51.71}$ & $\underset{\left(209\right)}{51.69}$ &  & $\left[51.55,\,51.80\right]$\tabularnewline
		30 &  &  &  &  &  & $\underset{\left(61\right)}{59.12}$ & $\underset{\left(154\right)}{59.12}$ & $\underset{\left(266\right)}{59.16}$ &  & $\left[59.48,\,59.87\right]$\tabularnewline
		50 &  &  &  &  &  & $\underset{\left(123\right)}{69.38}$ & $\underset{\left(225\right)}{69.36}$ & $\underset{\left(424\right)}{69.49}$ &  & $\left[69.56,\,69.95\right]$\tabularnewline
		100 &  &  &  &  &  & $\underset{\left(164\right)}{82.99}$ & $\underset{\left(400\right)}{83.26}$ & $\underset{\left(733\right)}{83.35}$ &  & $\left[83.36,\,83.86\right]$\tabularnewline
		\bottomrule
	\end{tabular}
		\par\end{centering}
	\caption{\label{tab:GPR_TE_MAX_CV}American price results for a Call on the Maximum option by using the
		GPR-Tree and GPR-EI methods (with control variate technique). In the
		last column the exact benchmark. The values in brackets are the computational
		times (in seconds).}
\end{table}
 \FloatBarrier
\subsection{Variance Reduction}

We conclude our numerical investigations by showing the effect of introducing a control variate on the variance of the estimated prices. 
In particular, we consider the same Geometric Put option as in Section \ref{PUT} and we price the same option $100$ different times, changing the seed of the Monte Carlo generator. This allows us to estimate the variance of the price estimator and to make comparisons. 
Results are available in Tables \ref{tab:STD-LMC} and \ref{tab:STD-GPR-MC}, that report the estimated standard deviations and their $95\%$ confidence intervals, computed according to the method suggested by Sheskin \cite{sheskin2000parametric}. It is evident that the the standard deviation (and thus the variance) of the prices obtained with the GPR-MC-CV method is several time lower than the one computed with the GPR-MC method. This is also confirmed for all the considered  combination of $P$ and $M$, by the Hartley's $F_{max}$ test (see Sheskin \cite{sheskin2000parametric}) with a $99\%$ confidence level.

\begin{table}
	\begin{centering}
 \begin{tabular}{cccccccccc}
 	\toprule 
 	& $P$ & \multicolumn{2}{c}{$250$} &  & \multicolumn{2}{c}{$500$} &  & \multicolumn{2}{c}{$1000$}\tabularnewline
 	\cmidrule{2-4} \cmidrule{6-7} \cmidrule{9-10} 
 	$d$ & $M$ & $10^{3}$ & $10^{4}$ &  & $10^{3}$ & $10^{4}$ &  & $10^{3}$ & $10^{4}$\tabularnewline
 	\cmidrule{2-10} 
 	2 &  & $\underset{\left[61.3,81.1\right]}{69.8}$ & $\underset{\left[19.6,25.9\right]}{22.3}$ &  & $\underset{\left[57.6,76.3\right]}{65.7}$ & $\underset{\left[18.0,23.8\right]}{20.5}$ &  & $\underset{\left[56.3,74.5\right]}{64.2}$ & $\underset{\left[17.7,23.4\right]}{20.2}$\tabularnewline
 	5 &  & $\underset{\left[47.1,62.3\right]}{53.6}$ & $\underset{\left[15.1,20.0\right]}{17.2}$ &  & $\underset{\left[48.5,64.2\right]}{55.2}$ & $\underset{\left[16.2,21.4\right]}{18.5}$ &  & $\underset{\left[41.3,54.6\right]}{47.0}$ & $\underset{\left[12.7,16.9\right]}{14.5}$\tabularnewline
 	10 &  & $\underset{\left[46.7,61.8\right]}{53.2}$ & $\underset{\left[14.1,18.7\right]}{16.1}$ &  & $\underset{\left[44.0,58.3\right]}{50.1}$ & $\underset{\left[13.2,17.4\right]}{15.0}$ &  & $\underset{\left[42.8,56.6\right]}{48.7}$ & $\underset{\left[13.1,17.3\right]}{14.9}$\tabularnewline
 	20 &  & $\underset{\left[46.6,61.7\right]}{53.1}$ & $\underset{\left[14.7,19.5\right]}{16.8}$ &  & $\underset{\left[43.8,58.0\right]}{49.9}$ & $\underset{\left[14.4,19.1\right]}{16.4}$ &  & $\underset{\left[40.8,53.9\right]}{46.4}$ & $\underset{\left[13.3,17.7\right]}{15.2}$\tabularnewline
 	40 &  & $\underset{\left[66.2,87.6\right]}{75.4}$ & $\underset{\left[21.6,28.5\right]}{24.6}$ &  & $\underset{\left[150.1,66.3\right]}{57.1}$ & $\underset{\left[15.1,19.9\right]}{17.2}$ &  & $\underset{\left[44.0,58.2\right]}{50.1}$ & $\underset{\left[13.3,17.6\right]}{15.1}$\tabularnewline
 	100 &  & $\underset{\left[6.75,89.3\right]}{76.8}$ & $\underset{\left[21.4,28.3\right]}{24.3}$ &  & $\underset{\left[60.0,79.4\right]}{68.3}$ & $\underset{\left[18.0,23.8\right]}{20.6}$ &  & $\underset{\left[53.6,70.9\right]}{61.1}$ & $\underset{\left[16.7,23.2\right]}{19.0}$\tabularnewline
 	\bottomrule
 \end{tabular}
		\par\end{centering}
	\caption{\label{tab:STD-LMC} Standard deviation for the prices of an American Geometric Basket Put option computed by means of the GPR-MC method (100 repetitions). Values between
		brackets are $95\%$ confidence intervals for the standard deviation. All results must be multiplied		by $10^{-3}$.}
 	
\vspace{7mm} 

	\begin{centering}
 \begin{tabular}{cccccccccc}
 	\toprule 
 	& $P$ & \multicolumn{2}{c}{$250$} &  & \multicolumn{2}{c}{$500$} &  & \multicolumn{2}{c}{$1000$}\tabularnewline
 	\cmidrule{2-4} \cmidrule{6-7} \cmidrule{9-10} 
 	$d$ & $M$ & $10^{3}$ & $10^{4}$ &  & $10^{3}$ & $10^{4}$ &  & $10^{3}$ & $10^{4}$\tabularnewline
 	\cmidrule{2-10} 
 	2 &  & $\underset{\left[8.0,10.5\right]}{9.1}$ & $\underset{\left[2.4,3.2\right]}{2.7}$ &  & $\underset{\left[7.7,10.2\right]}{8.8}$ & $\underset{\left[2.3,3.1\right]}{2.6}$ &  & $\underset{\left[7.4,9.8\right]}{8.5}$ & $\underset{\left[2.2,2.9\right]}{2.5}$\tabularnewline
 	5 &  & $\underset{\left[6.6,8.8\right]}{7.5}$ & $\underset{\left[2.4,3.1\right]}{2.7}$ &  & $\underset{\left[7.4,9.8\right]}{8.4}$ & $\underset{\left[2.3,3.1\right]}{2.7}$ &  & $\underset{\left[7.7,10.1\right]}{8.7}$ & $\underset{\left[2.0,2.7\right]}{2.3}$\tabularnewline
 	10 &  & $\underset{\left[6.7,8.9\right]}{7.7}$ & $\underset{\left[2.1,2.8\right]}{2.4}$ &  & $\underset{\left[7.1,9.4\right]}{8.1}$ & $\underset{\left[2.3,3.1\right]}{2.6}$ &  & $\underset{\left[6.3,8.3\right]}{7.1}$ & $\underset{\left[2.4,3.2\right]}{2.7}$\tabularnewline
 	20 &  & $\underset{\left[7.1,9.3\right]}{8.0}$ & $\underset{\left[2.4,3.1\right]}{2.7}$ &  & $\underset{\left[6.2,8.2\right]}{7.1}$ & $\underset{\left[2.9,3.8\right]}{3.3}$ &  & $\underset{\left[6.1,8.1\right]}{6.9}$ & $\underset{\left[2.0,2.7\right]}{2.3}$\tabularnewline
 	40 &  & $\underset{\left[6.7,8.9\right]}{7.6}$ & $\underset{\left[2.0,2.7\right]}{2.3}$ &  & $\underset{\left[7.7,10.2\right]}{8.7}$ & $\underset{\left[2.1,2.8\right]}{2.4}$ &  & $\underset{\left[7.3,9.7\right]}{8.4}$ & $\underset{\left[3.0,3.9\right]}{3.4}$\tabularnewline
 	100 &  & $\underset{\left[6.5,8.6\right]}{7.4}$ & $\underset{\left[2.9,3.8\right]}{3.2}$ &  & $\underset{\left[7.3,9.6\right]}{8.3}$ & $\underset{\left[2.3,3.1\right]}{2.7}$ &  & $\underset{\left[9.3,12.2\right]}{10.5}$ & $\underset{\left[2.7,3.5\right]}{3.0}$\tabularnewline
 	\bottomrule
 \end{tabular}
		\par\end{centering}
	\caption{\label{tab:STD-GPR-MC} Standard deviation for the prices of an American Geometric Basket Put option computed by means of the GPR-MC-CV method (100 repetitions). Values between
		brackets are $95\%$ confidence intervals for the standard deviation. All results must be multiplied		by $10^{-3}$.}
\end{table}
\FloatBarrier
	\section{Conclusions}
In this paper we have proposed  a new approach   to price American options on baskets of assets, each of them following a Black-Scholes dynamics. The method employs Machine Learning technique,  Monte Carlo method and  variance reduction technique that exploits the European option price as a control variate. The European prices are computed by means of a semy-analitical formula or Quasi-Monte Carlo simulations. Numerical results show that the method is reliable and fast for baskets including up to $100$ assets. The use of a control variate improves the algorithm accuracy and reduces the variance of the estimated prices. In certain cases, also the GPR-Tree and GRP-EI methods benefit from the use of a control variate. The computation time is small and shortly growing with respect to the dimension of the basket. Moreover,  the algorithm is partially parallelizable and therefore the computing time can be significantly reduced. Machine Learning seems to be a very promising tool for American option pricing in high dimension, overcoming the problem of the curse of dimensionality.

	\bibliographystyle{acm}
	\bibliography{bibliography}
	
\end{document}